\documentstyle[eqsecnum,array,multicol,prb,aps,psfig]{revtex}
\draft
\tighten
\begin{document}

\title{Charging Spectrum and Configurations
of a Wigner Crystal Island.}
\author{A.~ A.~ Koulakov and B.~ I.~ Shklovskii}
\address{Theoretical Physics Institute, University of Minnesota,
Minneapolis, Minnesota 55455}

\date{\today}

\maketitle
\begin{abstract}

Charging of a clean two-dimensional island is studied in the regime of small
concentration of electrons when they form the Wigner crystal. 
Two forms of electron-electron interaction potential are studied:
the pure Coulomb interaction and the exponential interaction corresponding
to the screening by a pair of close metallic gates.
The electrons are assumed to reside in a parabolic external confining potential.
Due to the crystalline symmetry the center of the confinement can be situated
at distinct positions with respect to the crystal. 
With the increasing number of electrons $N$ the center periodically
hops from one such a location to another providing the lowest total energy.
These events occur with the period $\sim N^{1/2}$.
At these moments in the case of the pure Coulomb interaction
the charging energy of the island has a negative correction $\approx 15\%$.
For the case of the exponential interaction at the moments of switching
the capacitance becomes negative and
$\sim N^{1/4}$ new electrons enter the island simultaneously.
The configurations of disclinations and dislocations in the island are also studied.
\end{abstract}
\begin{multicols}{2}

%
%
%
%

\section{Introduction}
\label{Introduction}

In recent experiments\cite{AshooriNew,AshooriOld}
the charging of a quantum dot is studied by the single
electron capacitance spectroscopy method. The quantum dot is located
between two capacitor plates: a metallic gate and a heavily doped GaAs layer.
Tunneling between the dot and the heavily doped side is possible during
the experimental times while the barrier to the metal is completely insulating.
DC potential $V_g$ and a weak AC potential are applied to the
capacitor. With the increase of $V_g$ the differential capacitance experiences
periodic peaks when addition of a new electron to the dot  becomes possible.
The spacing between two nearest peaks $\Delta V_{g}$ can be related to the ground state
energy $E(N)$ of the dot with $N$ electrons:
\begin{equation}
\begin{array}{ll}
{\displaystyle
\alpha e \Delta V_{g}}&{\displaystyle=E(N+1) - 2E(N) + E(N-1)
} \\ \\
{}&{\displaystyle =\Delta(N) \equiv e^{2}/C_{N}.}
\end{array}
\label{capacitance}
\end{equation}
Here $\alpha$ is a geometrical coefficient, 
$\Delta(N)$ is the charging energy, $C_{N}$ is
the capacitance of the dot with $N$ electrons. 
It was observed in Refs.~\onlinecite{AshooriNew,AshooriOld}
that at a low concentration of electrons or in a strong magnetic field
the nearest peaks can merge, indicating that at some values of $V_{g}$ 
two or even three electrons enter the dot simultaneously. 
In other words some charging energies apparently become zero or negative.
In a fixed magnetic field this puzzling event
repeats periodically in $N$. Disappearance of the charging energy looks
like a result of an unknown attraction between electrons and represents a
real challenge for theory.

Pairing of the differential capacitance peaks has been studied theoretically 
before for {\em disordered} dots.
Explanation of the pairing based on the 
{\em lattice} polaronic mechanism has been suggested in  
Ref.~\onlinecite{Phillips}. 
In Ref.~\onlinecite{Raikh} it was demonstrated how
electron-electron repulsion, screened by a close metallic gate,
can lead to electron pairing 
for a specially arranged compact clusters of localized states
in a disordered dot.
This effect is a result of redistribution of the other electrons after 
arrival of new ones. It was interpreted in Ref.~\onlinecite{Raikh}
as {\em electronic bipolaron}.

In this paper we study the addition spectrum of a dot in which the density of 
electrons is small and the external disorder potential is very weak,
so that electrons in the dot form the Wigner crystal.  
We call such a dot a Wigner crystal island. 
In the experimental conditions of 
Ref.\onlinecite{AshooriNew} 
one can think about a Wigner crystal island literally
only in the highest magnetic field. One can also imagine
similar experiments with a Wigner crystal island on the surface of liquid helium.
In the present paper we consider the extreme classical limit of the Wigner crystal,
when the amplitude of the quantum fluctuations is much smaller than
the interparticle distance. In this case one can think of electrons
as of classical particles and the energy of the system is given
by the following expression:  
\begin{equation}
E = \sum_{i < j} {U\left( \bbox{r}_i - \bbox{r}_j \right) } + A \sum_i \bbox{r}_i^2.
\label{Energy}
\end{equation}
The first term represents interactions among electrons located at points $\bbox{r}_i$,
with $U\left( \bbox{r}\right)$ being the interaction potential.
The second term is the contribution to the energy due to the external
confinement, which is assumed to have parabolic form.
Coefficient $A$ plays the role of strength of the confinement.
The forms of the interaction potential considered are:
the pure Coulomb interaction
\begin{equation}
U\left( \bbox{r}\right) = {e^2} / {\kappa r},
\label{Coulomb_int}
\end{equation}
with $e$ and $\kappa$ being
the electron charge and the dielectric constant correspondingly,
and the exponential interaction
\begin{equation}
U\left( \bbox{r}\right) = U_0\exp \left( -r/s \right).
\label{Exp_int}
\end{equation}
The latter interaction potential corresponds to the case
when the island is situated between two metallic gates,
with $s$ being of the order of the distance between them and
$U_0 \sim e^2/\kappa s$. In this paper we study the case $a\gg s$, where
$a$ is the average interparticle distance.

First we study the addition spectrum $E(N)$ of such a system numerically. 
We notice that for the both types of interaction the energy of the ground state has a
quasi-periodic correction. The period of this correction
scales as $\sim \sqrt{N}$, or the number of crystalline rows in the island.
Its shape is universal, i.e. independent of the form of 
the electron-electron interaction. Explanation of these oscillations is
the subject of the subsequent theoretical analysis.

We attribute the aforementioned oscillations to the combination of two effects: 
insertion of new crystalline rows and
motion of the center of the confinement relative to the crystal. 
Consider the former effect first.
Let us fix the position of the center of the external parabola
relative to the adjacent Bravais unit cell. 
For example let it coincide with one of the lattice sites.
As electrons are added to such a system, the number
of crystalline rows grows roughly as $\sim \sqrt{N}$.
The periodic appearances of the new rows bring about oscillations 
of the total energy $E(N)$ with $N$. The period of
these oscillations scales as
$\delta N \sim \sqrt{N}$ in agreement with our numerical results. 

Let us discuss the influence of the position of the confinement center.
The energy of the system 
as a function of this position has the same symmetry 
as the lattice. Hence the extrema of the energy have to be situated at the
points of high symmetry, e.g. the centers of 
two-fold rotational symmetry or higher. 
There are three such points in the two-dimensional (2D) 
triangular lattice (see Fig.~\ref{fig60}a below).  
The evolution of the island with $N$ 
consists in switching between these three positions
of the center, every time choosing the location providing the lowest total energy.
This effect is only based on the symmetry considerations and is therefore
universal, i.e. independent of the form of interaction.
This idea suggests a simple recipe for the calculation 
of the energy of the island.
One has to calculate three energy branches
corresponding to the different locations of the center and then choose
the lowest one. 

\begin{figure}
\centerline{
\psfig{file=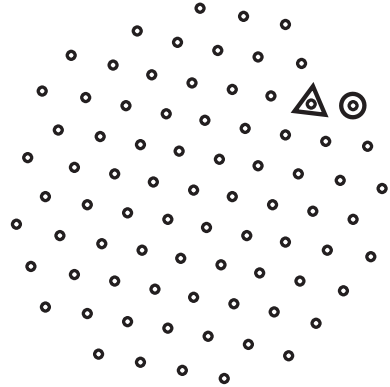,height=2.0in,bbllx=225pt,bblly=441pt,bburx=400pt,bbury=618pt}
}
\vspace{0.1in} 
\setlength{\columnwidth}{3.2in}
\centerline{\caption{
The configuration of electrons in the island for the case 
\protect{$N=80, A=10^{-8}, U_0=1$, and $s=1$}. 
A typical surface particle in a large dot has four nearest neighbors.
Two particles with three and five nearest neighbors are shown by the circle
and the triangle correspondingly.
They represent a pair of opposite surface disclinations or a surface dislocation.
\label{fig10}
}}
\vspace{-0.1in}
\end{figure}

Let us consider these two effects in more detail,
separately for the two types of interaction.
We start with the case of extremely short-range interaction 
given by Eq.~(\ref{Exp_int}): $a\gg s$. 
This case can be realized, e.~g., if the confining 
parabola is very weak compared to the interaction prefactor: 
$AR^2 \ll U_0$.
Here $R$ is the radius of the island.
For this case the interaction between the nearest neighbors
is a very steep function of distance due to the fast decay 
of the exponential (\ref{Exp_int}).
The variations of the lattice constant in this case are very small
(see a more elaborate discussion in Section~\ref{Inc_Isl_ATheory}).
This implies that the crystalline rows are almost straight lines
(see Fig.~\ref{fig10}).
Hence we deal with a piece of almost perfect triangular crystal with new
electrons being added on the surface of the island. A new crystalline row
therefore appears on the surface.
It can be shown that this creates an anomalous increase in the
density of states (DOS) of electrons. Such variations of DOS result in
the appearance of the periodic correction to the energy of the island.
This correction is considered in detail in Section~\ref{Inc_Isl}.
An interesting implication of this picture is the multiple electron
entering. It turns out that 
if one slowly raises the chemical potential of electrons in the island,
then at the points of switching between the branches
mentioned above about $N^{1/4}$ electrons enter the island simultaneously.
A simple model of this phenomenon has been suggested before in 
Ref.~\onlinecite{Us_Phyl_Mag} for a small island containing $\lesssim 55$ electrons.
In this paper we discuss this phenomenon for larger islands.

%
%

One can think about the new rows appearing in the crystalline island
as of pairs of dislocations of opposite sign. 
In the case of the short-range
interaction these defects are pushed to the surface by a huge
price for elastic deformations. 
In the Coulomb case, described by Eq.~(\ref{Coulomb_int}), the 
shear modulus and, subsequently, the
Young's modulus of the crystal are relatively small.
As discussed in Section~\ref{Defects} 
the incommensurability of the circular shape of the dot 
with the lattice and the inhomogeneity of the density of electrons 
generate in this case topological defects, disclinations and dislocations,
{\em inside} the island. 
We argue that these defects determine the variations of the energy
when the center of the confinement is fixed relative to the crystal.
The number of dislocations scales as the number of crystalline rows $\sim \sqrt{N}$.
This implies that a new dislocation appears every $\sim \sqrt{N}$ electrons.
Due to the discreteness of these defects 
a single branch of the total energy 
corresponding to a fixed position of the center
acquires a quasi-periodic correction.
This periodic building up and relaxation of the elastic energy
of the dot caused by the discreteness of dislocations is very similar 
to the variations of the electrostatic energy brought about by the discreteness
of electrons, known as Coulomb blockade. 
Following this analogy we call the former periodic phenomenon
an elastic blockade.

The elastic blockade appears to be a bit more complicated than 
its electrostatic counterpart. 
The center of the confinement can move 
among three distinct points of the triangular lattice
mentioned above making the system
switch from one energy branch to another. 
Such a sudden switching results in 
a correction to the charging energy [see Eq.~(\ref{capacitance})]:
\begin{equation}
\delta \Delta \approx -0.15 \overline{\Delta},
\end{equation}
where $\overline{\Delta}$ is the average charging energy.
This reduction of the distance between the nearest Coulomb blockade
peaks happens with the period $\sim \sqrt{N}$ determined by the elastic 
blockade. It can be thought of as an analog of merging of a few peaks 
observed in the case of the short-range interaction.

The optimum configurations of electrons in the parabolic confinement with the
Coulomb interaction have been studied before in Ref.~\onlinecite{Peeters94}.
Our numerical results both for the energies and the configurations
agree with the results of this work. However our interpretation
of the results is different.
The authors of Ref.~\onlinecite{Peeters94} adopt a model in which the
electrons fill in shells concentric to the perimeter of the island.  
We argue that in the regime studied in the numerical experiment
only a narrow ring adjacent to the perimeter is concentric to it.
The width of such a ring is $\sim \sqrt{R\cdot a}$.
The rest of the island is filled with an almost perfect crystal. 
(see a Sections~\ref{Defects} and \ref{Compr_Isl}).

The paper is organized as follows. First we consider the case of an extremely
short-range interaction [Eq.~(\ref{Exp_int})], when the Young's modulus of the crystal is very
large and no lattice defects can exist in the interior of the island. 
This case is discussed in Sections~\ref{Inc_Isl} and \ref{Inc_Isl_ATheory}.
In Section~\ref{Coulomb} we report the results of the numerical 
solution of the problem with the Coulomb interaction [Eq.~(\ref{Coulomb_int})].
In Section~\ref{Defects} we turn to the discussion of different kinds of defects,
which can exist in a compressible island.
In Section~\ref{Compr_Isl}, 
we discuss the theory of the elastic blockade for the case of the Coulomb interaction.
Section~\ref{Conclusions} is dedicated to our conclusions.

%
%
%
%

\section{Short-Range interaction: Numerical Results}
\label{Inc_Isl}

In this section we consider the system of $N$ electrons interacting by
a strongly screened short-range potential. This limit is 
defined by Eq.~(\ref{Energy}) with interaction given by Eq.~(\ref{Exp_int})
and $a\gg s$. 
As it is shown later in Section~\ref{Inc_Isl_ATheory}
the latter condition can be realized if the interaction prefactor 
significantly exceeds the typical confinement energy: $U_0 \gg AR^2$. 
At this condition the variations of the lattice constant of the crystal 
are negligible. This indeed can be observed 
in the configurations obtained in the numerical experiment.
One of such a configurations is shown in Fig.\ref{fig10},
obtained at $a \approx 13 s$.

In our numerical analysis we minimized the energy functional (\ref{Energy}) with
the exponential interaction given by Eq.~(\ref{Exp_int}) 
with respect to the coordinates of electrons using
a genetic algorithm similar to that outlined in Ref.~\onlinecite{Morris96}.
Below we describe this numerical technique 
and the motivation for its use.

The problem at hand belongs to the vast class of problems 
of finding the global minimum of a multi-dimensional function,
which has plenty of local minima. The well developed methods for
convex differentiable functions (the conjugate gradient, Newton's method) do not work
here as they find only some local minimum. The most frequently used method
in this case is the Metropolis simulated annealing technique.~\cite{Peeters94}
In this method the system is modeled at some artificially introduced
temperature, which is gradually decreased to a very small value.
It is assumed that after this annealing the system falls into the
state of the lowest energy.
Although in principle
in this method the system can hop from a metastable state to the ground 
state, in practice, if the potential barrier between them is high,
the time to perform such a hop can exceed the time of the simulation.

An alternative method is the genetic algorithm, which was proved to be 
superior to the simulated annealing.\cite{Morris96}
Initially five different parent configurations were obtained by
relaxing random configurations using the conjugate gradient algorithm.
Then these configuration were mated pairwise to obtain additional
$15$ child configurations (including mating with itself), 
which were again relaxed using the conjugate gradient
algorithm. Mating consisted in cutting two configurations into halves
by a random line and then connecting those halves belonging to different
parents to form a new child. From the resulting $20$ conformations (parents and
children) five were chosen to be parents for the next iteration.
The new parents consisted of the lowest energy configuration and
the other four separated by at least the precision of the conjugate gradient.
This has been done to avoid dominating the process by one conformation.
During ten iterations as many as $155$ local minima were examined.
The energies of the optimum conformations obtained in this way
agree with those of Ref.~\onlinecite{Peeters94}
and for some $N$ are even lower. 
Before presenting these results  
we would like to discuss the method we used to process the data.

\begin{figure}
\centerline{
\psfig{file=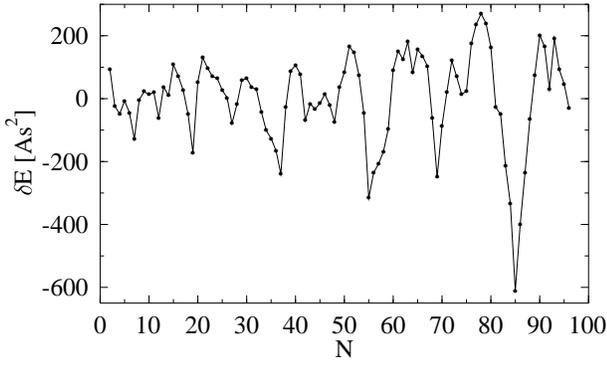,height=1.9in,bbllx=60pt,bblly=103pt,bburx=489pt,bbury=369pt}
}
\vspace{0.1in} 
\setlength{\columnwidth}{3.2in}
\centerline{\caption{
The fluctuating part of the ground state energy in the model with
the exponential interaction (\protect{\ref{Exp_int}}) 
for $A=10^{-8}, U_0=1$, and $s=1$.
\label{fig20}
}}
\vspace{-0.1in}
\end{figure}

To analyze the dependence of the ground state energy $E \left( N\right)$ on
the number of particles we split it into the smooth $\bar{E} \left( N\right)$ 
and the fluctuating component $\delta E \left( N\right)$: 
\begin{equation}
E \left( N\right) = \bar{E} \left( N\right) + \delta E \left( N\right),
\label{dE}
\end{equation}
in the manner of Ref.~\onlinecite{Us_Phyl_Mag}.
The smooth component has the form
$\bar{E}(N) = \eta_1 N^2 + \eta_2 N + \eta_3 N^{2/3} + \eta_4 N^{1/2} + \eta_5 N^{1/3}$,
where the coefficients $\eta_1 \cdots \eta_5$ 
are chosen to minimize the fluctuations.
The fluctuating part is displayed in Fig.~\ref{fig20}.

%
%

The curve in Fig.~\ref{fig20} evidently has a quasi-periodic structure.
It consists of the series of interchanging deep and shallow 
minima, separated by the peeks with more or less 
smooth slopes.
We call this structure the "Kremlin wall" for its resemblance to the
embattlements on top of the walls of the Moscow Kremlin.
Its period with a very good precision scales as $N^{1/2}$ and
numerically is equal to the number of electrons in the outer
crystalline row of the island. 
The amplitude of the oscillations grows 
with $N$ as~\cite{Us_Phyl_Mag}
\begin{equation}
\left| \delta E \right| \propto N^\gamma, ~~ \gamma = 0.8 \pm 0.1.
\label{Scaling}
\end{equation}

Although the crystal in the bulk is almost perfect, its boundary is extremely 
irregular (see Fig.~\ref{fig10}). 
It can be thought of as a superposition of various types of
defects pushed against the surface by extremely large Young's modulus,
provided by the short-range interaction. 
These defects can be associated with the particles having an anomalous
coordination number. Normally a particle on the surface of the triangular
crystal has four nearest neighbors. But there are particles having
a coordination number equal to three or five. These particles can be associated
with positive and negative disclinations on the surface correspondingly, 
as their creation assumes removal (insertion) of a $\pi/3$ wedge (see Fig.~\ref{fig10}).
Dislocations on the surface are pairs of such disclinations of the opposite sign
forming a dipole.
At a small Young's
modulus however these defects can dive inside the island. This transition
is quantitatively described in Section~\ref{Defects}.

%
%
%
%

\section{Short-Range interaction: a Theory}
\label{Inc_Isl_ATheory}

In this section we use a hard disk model to explain our
numerical results obtained in the limit $U_0 \gg AR^2$. 
To justify this model we first show that the variation of the lattice constant  
in the island is small in this limit:
\begin{equation}
\delta a = a(R) - a(0) = \frac s2 \ln N \ll a(0).
\end{equation} 
Then we demonstrate that the interaction energy is small compared to
the confinement energy and hence can be neglected.

To prove the first statement we find the pressure $\sigma_{\rm rr}$
in the crystal associated with its contraction by the
external potential $Ar^2$. The solution can be found similar to 
Ref.~\onlinecite{LL7_1}
\begin{equation}
\sigma _{\rm rr} =  - S(\sigma) A(R^2-r^2),
\label{sigma_rr}
\end{equation}
where $S(\sigma) = (3+\sigma)/4$, $3/4 \le S(\sigma) \le 1$, 
and $\sigma$ is the Poisson ratio.
The solution is easy to understand for the limiting case of liquid 
$\sigma \rightarrow 1$, when $\sigma_{ik} = -p \delta_{ik}$, $p$ being pressure.
In this case $p$ is of purely hydrostatic origin:
\begin{equation}
\frac {dp(r)} {dr} = - 2Arn(r).
\end{equation}
Assuming the density $n(r)$ to be uniform 
and $p(R)=0$ we obtain the following solution:
\begin{equation}
p (r) = An(R^2-r^2),
\end{equation}
which agrees with Eq.~(\ref{sigma_rr}) taken at $\sigma = 1$. 
The forces produced by this pressure have to be balanced by the
interaction forces between particles:
\begin{equation}
f \sim \sigma _{\rm rr}a \sim An(R^2-r^2)a \sim \frac {U_0}s e^{-a/s}.
\label{force}
\end{equation}
In this equation we assumed $S(\sigma ) \sim 1$.
From this condition we obtain the following result for
the lattice spacing in the island:
\begin{equation}
a(r) \approx s \ln \left[ \frac {U_0}{A \left(R^2-r^2 \right)} \right].
\label{a_r}
\end{equation}
The difference between the lattice spacing on the boundary and in the center
can be readily obtained:
\begin{equation}
\delta a \approx s \ln \left( \frac {U_0}{ARa} \right) -
s \ln \left( \frac {U_0}{AR^2} \right) = s\ln \frac Ra = \frac s2 \ln N.
\label{delta_a}
\end{equation}
Comparing (\ref{a_r}) and (\ref{delta_a}) one can see that the case
$a \gg \delta a$ can be realized when $U_0 \gg AR^2$. This implies that
the prefactor of the interaction potential has to be large. In this case
the nearest particles are far away from each other where the exponential 
interaction is very steep. This eliminates significant changes
in the interparticle distance. 

Although the derivative of the interaction potential 
is important in determining the lattice spacing
the typical value of interaction energy in this case is very small.
The characteristic interaction energy per particle can be 
estimated as: $E_{\rm nn} \sim U_0 \exp (-a/s)$. 
From Eq.~(\ref{force}) we conclude that
\begin{equation}
U_0e^{-a/s} \lesssim \frac sa AR^2.
\end{equation}
Hence in the considered regime $a \gg s$ the interaction energy 
indeed can be neglected. 

It is plausible then to accept the hard disk
model to explain our numerical results obtained for this case. 
We assume therefore that the interaction has the form:
\begin{equation}
U(r)=\left\{
\begin{array}{ll}
{\displaystyle
	\infty, }&{r<a,
} \\ \\
{\displaystyle
	0,} &{r \ge a.
}
\end{array}
\right.
\end{equation}
In this model the interaction energy is zero
and the total energy of the system can be written as:
\begin{equation}
E=A\sum_{i=1}^N {\bbox r}_i^2,
\label{Hard_Disk_Energy}
\end{equation}
where ${\bbox r}_i$ belong to the triangular lattice, with the lattice
spacing equal to the radius of the interaction $a$.
This energy formally coincides with the moment of inertia of the system
of $N$ particles of mass $A$. Hence to find the minimum energy configuration
one has to cut a piece from the triangular crystal that has a minimum moment of inertia.
This piece must contain the given number of particles $N$.
The average values of the total energy, chemical potential and
the charging energy are given by:
\begin{equation}
\begin{array}{l}
{\bar{E} = AN\cdot R^2/2 = AN^2/2\pi n} \\ \\
{\bar{\mu} = d\bar{E}/dN = AN/\pi n} \\ \\
{\bar{\Delta} = d\bar{\mu}/dN = A/\pi n},
\end{array}
\label{Averages}
\end{equation}
where $R$ is the average radius of the circle, $n\pi R^2=N$,
$n = 2/a^2\sqrt{3}$ is the concentration of lattice sites.

A similar problem has been considered in the literature 
in the context of the so-called circle problem. 
The problem is to 
find the fluctuations of the number of particles 
belonging to some lattice inside of a circle of radius $R$ 
(see Ref.~\onlinecite{Dyson}). 
The quantity which has been studied is the deviation of the
number of particles from its average value:
\begin{equation} 
\delta N(R) \equiv N(R)-n\pi R^2,
\end{equation}
where $n\pi R^2 = \bar{N}$ is the average number of particles in the circle.
The classical circle problem which goes back to Gauss is to
find the uniform bound for $|\delta N(R)|$.
The best result in this direction is\cite{Dyson}
\begin{equation}
|\delta N| \le C_{\epsilon} R^{46/73+\epsilon} \sim \bar{N} ^{23/73+\epsilon /2}
\label{Upper_Bound}
\end{equation}
Here $C_{\epsilon}$ is a $R$-independent constant and $\epsilon > 0$.
A simple estimate for these fluctuations can be obtained from the
assumption that they occur around the perimeter of the circle~\cite{Dyson}
in the strip of width of the order of the lattice spacing $a$.
The number of such points $N_{\rm per} \sim R/a$ and the fluctuations are
\begin{equation}
\delta N \sim \sqrt{N_{\rm per}} \sim \sqrt{R/a} \sim \bar{N}^{1/4}.
\label{Delta_N_ave}
\end{equation}
This estimate may imply a pseudo-random behavior of $\delta N$.
Notice that it agrees with the uniform upper bound (\ref{Upper_Bound})
as $23/73 > 1/4$.

Ref.~\onlinecite{Dyson} also considers the distribution of $\delta N$. 
It was found that 
the distribution function is non-Gaussian:
\begin{equation}
p(\delta N) \lesssim \exp \left( -\delta N^4/\bar{N} \right),
\label{Distribution}
\end{equation}
with the mean square deviation given by Eq.~(\ref{Delta_N_ave}).
It was also noted that $\delta N(R)$ in addition to be 
pseudo-random is almost periodic function of the radius.

%
%
\begin{figure}
\centerline{
\psfig{file=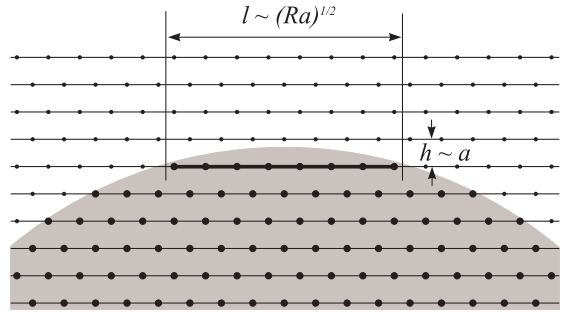,height=1.6in,bbllx=52pt,bblly=343pt,bburx=576pt,bbury=635pt}
}
\vspace{0.1in} 
\setlength{\columnwidth}{3.2in}
\centerline{\caption{
The circular island (gray) covering the crystalline rows, shown by lines.
The lattice points within the circle are shown by larger disks
than those outside. Every time the circle intersects a new row  
a new terrace appears. One of such terraces is shown by the bold segment.  
\label{fig30}
}}
\vspace{-0.1in}
\end{figure}

\begin{figure}
\centerline{
\psfig{file=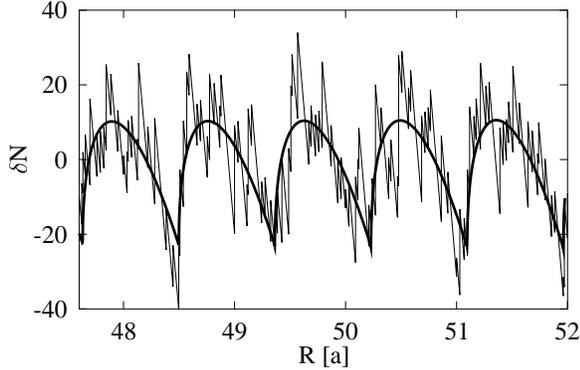,height=1.9in,bbllx=60pt,bblly=103pt,bburx=489pt,bbury=369pt}
}
\vspace{0.1in} 
\setlength{\columnwidth}{3.2in}
\centerline{\caption{
The averaged oscillations of the number of particles as a function of the radius
of the circle (bold smooth line) and the actual value, observed in the
numerical simulation on the lattice (thin rough line). 
The center of the circle is located at one of the lattice sites.
\label{fig40}
}}
\vspace{-0.1in}
\end{figure}

Below we will concentrate on the periodicity of
$\delta N(R)$, rather than on its randomness. In essence we would like
to find the average of $\delta N(R)$ over many periods
and understand the nature of its oscillations.
Then the question about the deviations from this average becomes
relevant.
Our analysis shows that the oscillations are associated with
the periodic intersections of the circle with the crystalline rows
(see Fig.~\ref{fig30}). At the moment of intersection an anomalously
large number of particles can be put into the circle. This brings
about an increase in the function $\delta N(R)$ every time the circle hits a new
crystalline row. This increase is accompanied by a decrease later in the period as
the average value of $\delta N$ is zero. To illustrate this point consider
the triangular lattice (Fig.~\ref{fig30}). In this case for the most part 
the oscillations of $\delta N$ are associated with the collisions of
the circle with the main crystalline rows, ($\sqrt{3},1$) plus other five
obtained by $\pi/3$ rotations, separated by $h=a\sqrt{3}/2$.
Hence the variations of $\delta N$ associated with these rows 
have a period in $R$ equal to $T_R=h$. One can collapse $\delta N$
from many such periods into one and average over all of them.
To do this we first define a normalized variation
of the number of particles:
\begin{equation}
\delta \eta(R) \equiv \frac {\delta N}{\sqrt{R}}
\label{Delta_Eta_Def}
\end{equation}
According to  Eq.~(\ref{Delta_N_ave}) 
the average amplitude of this function 
does not change with $R$. To extract the component of this function
having the period $T_R=h$ we average it over many such periods:
\begin{equation}
\overline{ \delta \eta }\left( R \right)  = \lim_{M\rightarrow \infty} 
\left[ 
\frac 1M \sum _{m=0}^M \delta \eta \left( T_Rm + R\right)
\right].
\label{Definition_of_Average}
\end{equation}
Thus obtained function 
$\overline{\delta N} (R) \equiv \overline{ \delta \eta }\left( R \right) \sqrt{R}$ 
is shown in Fig.~\ref{fig40}
by a smooth line. The actual $\delta N(R)$ is also presented in this
figure to show that its main variation is indeed associated with the 
aforementioned period. 
%
%

Before embarking into the calculation of the $\overline{\delta N}$
we would like to mention that other crystalline planes produce
similar oscillations. These oscillations have a period that is
smaller than $h$ and can be incommensurate to it. The
amplitude of these satellite oscillations is numerically smaller 
than the main one and hence they will be neglected in this paper.
All the periodicities added together produce a complex fractal
curve shown in Fig.~\ref{fig40}. 

Before getting into details of the averaging 
we would like to show a 
simple way to estimate the amplitude of these oscillations.  
Look at Fig.~\ref{fig30}. The collision of a crystalline row with
the circle results in the formation of the ``terrace'' on the 
surface of the crystal bounded by the circle. 
The average length of such a ``terrace'' is 
\begin{equation}
l\sim \sqrt{R\cdot a}.
\label{terrace_length}
\end{equation}
The number of crystalline sites
in such a ``terrace'' provides an estimate for the variations in the
number of particles inside the circle:
\begin{equation}
\delta N = l/a \sim \sqrt{R/a}.
\label{Another_Estimate}
\end{equation}
This estimate agrees with Eq. (\ref{Delta_N_ave}). However,
it is based on the existence of the ordered periodic structure
in $\delta N$.

This argument allows us to estimate the amplitude of the oscillations 
of the total energy (\ref{Hard_Disk_Energy}). Indeed, we have
found the oscillations of the number of particles as a function of 
radius or, in other words, of the chemical potential $\mu = AR^2$.
Now it easy to calculate the related oscillation of the chemical potential
as a function of the number of particles. To this end we notice that 
if the oscillations are weak
they are simply proportional to each other:
\begin{equation}
\delta \mu = - \delta N / \bar{\nu},
\label{Chem_Pot}
\end{equation} 
where $\bar{\nu}=1/\bar{\Delta}$ is the average density of states.
The period of $\delta \mu (N)$ is related to the
period of $\delta N(R)$ in the obvious way: 
$T_N=dN/dR \cdot T_R=\pi n Rh \sim \sqrt{N}$.
Now the oscillating part of the total energy can be easily
estimated:
\begin{equation}
\begin{array}{ll}
{\delta E(N)} & {= \int dN \delta \mu (N) } \\ \\ 
{} & {\sim T_N \cdot \delta N \bar{\Delta} \sim Aa^2 N^{3/4}}.
\end{array}
\label{Estimate_For_Delta_E}
\end{equation} 
Notice that both the periodicity and amplitude of the oscillations
given by this estimate agree with our
numerical results (\ref{Scaling}). 
Below we will calculate the form of these oscillations averaged over many
periods.

We now turn to the calculation of $\overline{\delta N} (R)$. 
We follow the guidelines of Ref.~\onlinecite{Dyson}.
Assume that the center of the circle is located at point
$\bbox{r}_0$ reduced to Bravais unit cell.
The number of particles in the circle can be 
expressed through the sum over the lattice points ${\bbox R}_l$:
\begin{equation}
\begin{array}{ll}
{\delta N(R)} & {= N - n\pi R^2 }\\ \\
{} & {\displaystyle   = \sum_{{\bbox R}_l} f\left( {\bbox R}_l \right) -
\int n d^2r f\left( {\bbox r}\right),
}
\end{array}
\label{Real_space_sum}
\end{equation}
where
\begin{equation}
f({\bbox r})=\left\{
\begin{array}{ll}
{\displaystyle
	1, }&{ \left| {\bbox r}-{\bbox r}_0 \right| \le R,
} \\ \\
{\displaystyle
	0,} &{ \left| {\bbox r}-{\bbox r}_0 \right| > R.
}
\end{array}
\label{Number_function}
\right.
\end{equation}
Using the Poisson summation formula, Eq.~(\ref{Real_space_sum})
can be rewritten as follows:
\begin{equation}
\delta N (R) = n \sum_{{\bbox Q}_i \neq 0} \tilde{f}\left( {\bbox Q}_i \right),
\end{equation}
where the summation is assumed over the reciprocal lattice vectors ${\bbox Q}_i$
and
$\tilde{f}\left( {\bbox q} \right)=2\pi R %
J_1 (\left| {\bbox q} \right|R)/\left| {\bbox q} \right| \cdot \exp (-i\bbox{qr}_0)$
is the Fourier transform of function (\ref{Number_function}), with
$J_1(x)$ being the Bessel function. For the normalized variation
of number of particles ~(\ref{Delta_Eta_Def}) we then obtain:
\begin{equation}
\delta \eta(R) =
2\pi n\sqrt{R} \sum_{{\bbox Q}_i \neq 0} 
\frac {J_1 (\left| {\bbox Q}_i \right|R)} {\left| {\bbox Q}_i \right|}
e^{-i\bbox{Q}_i \bbox{r}_0 }.
\label{Delta_Eta}
\end{equation}
The limit in Eq.~(\ref{Definition_of_Average})
can be most easily evaluated upon noticing that 
$J_1(x) \approx \sqrt{2/\pi x} \cos \left( x-3\pi/4 \right)$,
when $x \gg 1$. Changing the order of summation over $\bbox{Q}_i$ and $m$
we obtain:
\begin{equation}
\begin{array}{l}
{\displaystyle
\overline{ \delta \eta} \left( R \right) = 
\lim_{M\rightarrow \infty} \left[ \frac 1M 
2\pi n 
\sum_{{\bbox Q}_i \neq 0}
\sum _{m=0}^M
\sqrt{\frac 2{\pi}}
\right.
} \\ \\ 
{\displaystyle
\left.
\times
\frac {
\cos \left(
\left| {\bbox Q}_i \right| T_R m + \left| {\bbox Q}_i \right| R - 3\pi /4
\right)
}
{\left| {\bbox Q}_i \right|^{3/2}}
e^{-i\bbox{Q}_i \bbox{r}_0} \right]
}
\end{array}
\end{equation}
To evaluate the sum over $m$ we again can use the Poisson summation formula,
which establishes a selection rule for the values of
the wave-vector $\bbox{Q}_i$:
\begin{equation}
\left| {\bbox Q}_i \right|T_R = 2\pi n,
\label{Selection_Rule}
\end{equation}
where $n$ is an integer. The reciprocal lattice vectors are
representable in the form $\bbox{Q}_i = Q_0\left(k_1\sqrt{3}/2, k_1/2+k_2 \right)$,
where $Q_0=4\pi/a\sqrt{3}$ and $k_{1, 2}$ are integers.
Equation (\ref{Selection_Rule}) can be rewritten in term of $k_{1, 2}$ 
in the following way: $3k_1^2+(k_1 + 2k_2)^2=4n^2$, which is a Diophantine
equation. It has a trivial set of solutions, $k_1=0$, $k_2=n$ plus five others 
obtained from it by $\pi/3$ rotations. They correspond to the collisions of the 
circle with the main sequence of the crystalline rows, separated by $T_R$.
These solutions will be taken into account below. 
The other solutions of Eq.~(\ref{Selection_Rule}) are more sparse in the $Q$-space,
and hence produce smaller than $T_R$ periods in the real space. They
correspond to the collisions with other crystalline lines.
They produce numerically smaller variations than the main sequence and
will be disregarded below.
Thus we conclude that 
\begin{equation}
\begin{array}{l}
{\displaystyle
\lim_{M\rightarrow \infty} \left[
\frac 1M \sum_{m=0}^M 
\cos \left(
\left| {\bbox Q}_i \right| T_R m + \left| 
{\bbox Q}_i \right| R -	3\pi /4 \right) 
\right]} \\ \\
{\displaystyle
= \sum_{\bbox{\tilde{Q}}_i} \delta_{\bbox{\tilde{Q}}_i, \bbox{Q}_i}
\cos \left( \left| 
{\bbox Q}_i \right| R -	3\pi /4 \right)
,}
\end{array}
\end{equation}
where $\bbox{\tilde{Q}}_i$ are solutions of (\ref{Selection_Rule}) with the
reservations mentioned above. The remaining summation over $\bbox{Q}_i$
is easily performed and we obtain finally the result for the average variation
of the number of particles:
\begin{equation}
\overline{\delta N _{\bbox{r}_0}} \left( R \right)  = 
\sum_{i=1}^6 \overline{\delta N_0} \left(R - 
\bbox{\hat{e}}_i \bbox{r}_0
\right),
\label{Delta_N}
\end{equation}
where $\bbox{\hat{e}}_i$ are the unitary vectors normal to the
six main crystalline rows, $\bbox{\hat{e}}_i= \hat{U}^i (\sqrt{3}/2, 1/2)$,
$\hat{U}$ is the $\pi/3$ rotation, $i=0,\ldots,5$.
Function $\overline{\delta N_0 }$ determines the oscillations 
produced by one set of rows
if the center of the circle $\bbox{r}_0$ coincides with one of the
lattice sites:
\begin{equation}
\overline{\delta N_0 } \left(  R \right)=
\frac {2^{3/4} 3^{3/8}} {\pi ^{1/4}}
\left[ \bar{N} (R) \right] ^{1/4}
\zeta \left( -\frac 12, \frac {R}{T_R} \right).
\label{Delta_N_0}
\end{equation}
Here
\begin{equation}
\zeta \left( z, q \right) = \frac {2\Gamma (1-z)} {\left( 2\pi \right)^{1-z}}
\sum_{k=1}^{\infty} \frac {\sin \left( 2\pi k q + z\pi/2 \right)
} {k^{1-z}},
\label{Zeta_Function}
\end{equation}
is the generalized Riemann zeta-function.~\cite{Whittaker63}
In addition to the numerical coefficient in Eq.~(\ref{Delta_N_0}) we
obtained the overall amplitude $N^{1/4}$, which agrees with
our qualitative analysis, and the oscillating part 
$\zeta \left( -1/2, R/T_R \right)$, which has an amplitude of the order of unity.
In Fig.~\ref{fig40} $\overline{\delta N}_{\bbox{r}_0=0} $ is shown by a smooth line.
It is clear from Eq.~(\ref{Delta_N}) that six sequences of the crystalline rows 
contribute to oscillations independently, with $\bbox{\hat{e}}_i \bbox{r}_0$
determining a ``delay'' produced by the displacement of the
center of the circle from a lattice site.

We would like now to calculate the oscillations of the total energy
of the incompressible island. First, as parameters of the problem 
we will use $\bbox{r}_0$ and $R$. This implies that we will have a given
value of the chemical potential in the island $\mu=AR^2$. The relevant
thermodynamic potential in this case is $\Omega_{\bbox{r}_0}(\mu)$. 
We will evaluate this potential and
optimize it with respect to $r_0$.
We will then use the theorem of small increments to
find $E(N)$. This function will indeed be similar to the Kremlin wall 
structure displayed in Fig.~\ref{fig20}.

\begin{figure}
\centerline{
\psfig{file=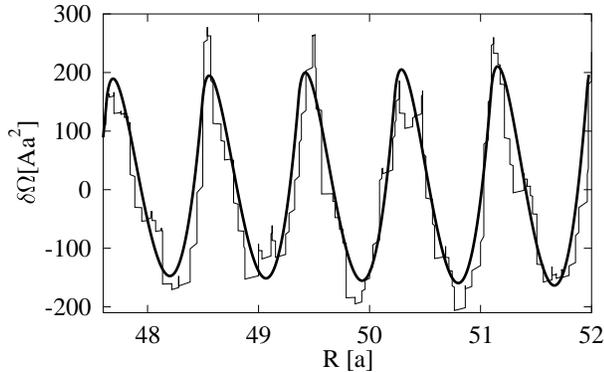,height=1.9in,bbllx=60pt,bblly=103pt,bburx=489pt,bbury=369pt}
}
\vspace{0.1in} 
\setlength{\columnwidth}{3.2in}
\centerline{\caption{
The averaged oscillations of the thermodynamic potential \protect{$\Omega$}
as a function of the radius of the circle (bold smooth line) and
the values actually observed in the numerical simulation (thin rough line).
The center of the circle is located at one of the lattice sites.
\label{fig50}
}}
\vspace{-0.1in}
\end{figure}

The first step in this program is realized similar 
to obtaining $\overline{\delta N}$. We apply the method used above
to the function
\begin{equation}
\Omega (R ) = E(R ) - \mu N, ~~\mu=AR^2.
\end{equation} 
The average value of this function is negative:
\begin{equation}
\bar{\Omega}=-AN\cdot R^2/2=-\bar{E},
\end{equation}
(compare to Eq.~(\ref{Averages})). The deviation of the potential from the average 
is given by
\begin{equation}
\begin{array}{ll}
{\delta \Omega (R)} & { = \delta E (R) - AR^2\delta N (R) } \\ \\ 
{} & {\displaystyle 
= \sum_{{\bbox R}_l} g\left( {\bbox R}_l \right) -
\int n d^2r g\left( {\bbox r}\right) - AR^2\delta N (R),
}
\end{array}
\label{Delta_Omega_Def}
\end{equation}
where 
\begin{equation}
g({\bbox r})=\left\{
\begin{array}{ll}
{\displaystyle
	A({\bbox r}-{\bbox r}_0)^2, }&{ \left| {\bbox r}-{\bbox r}_0 \right| \le R,
} \\ \\
{\displaystyle
	0,} &{ \left| {\bbox r}-{\bbox r}_0 \right| > R.
}
\end{array}
\label{Energy_function}
\right. 
\end{equation}
Using 
\begin{equation}
\tilde{g}\left( {\bbox q} \right)= \left[ \frac {2\pi A R^3}q
J_1 (\left| {\bbox q} \right|R)
- \frac {4\pi A R^2}{q^2}
J_2 (\left| {\bbox q} \right|R)
\right] e^{-i\bbox{qr}_0},
\end{equation}
one can obtain the oscillations of the thermodynamic potential in
exactly the same manner as those of the number of particles.
To this end, as it follows from (\ref{Estimate_For_Delta_E}),
one has to average the normalized variations of potential:
$\delta \omega \equiv \delta \Omega /R^{3/2}$ 
(compare to (\ref{Delta_Eta_Def})).
Eventually we obtain 
\begin{equation}
\overline{\delta \Omega _{\bbox{r}_0}} \left( R \right)  = 
\sum_{i=1}^6 \overline{\delta \Omega_0} \left( R - 
\bbox{\hat{e}}_i \bbox{r}_0
\right),
\label{Delta_Omega}
\end{equation}
where
\begin{equation}
\overline{\delta \Omega_0 } \left( R \right)= - \bar{\Delta}
\frac{ \pi^{1/4} 2^{9/4} } {3^{3/8}} 
\left[ \bar{N} ( R ) \right]^{3/4} \zeta \left( -\frac 32, \frac {R}{T_R} \right),
\label{Delta_Omega_0}
\end{equation}
is the function describing the oscillations if the center of the
circle coincides with one of the lattice sites. 
Its amplitude agrees with our presumption (\ref{Estimate_For_Delta_E}).
Notice that $\overline{\delta \Omega _{\bbox{r}_0}}$
given by (\ref{Delta_Omega}) is in fact 
a function of the chemical potential as the latter is related to the radius
by $\mu = AR^2$.
This function for $\bbox{r}_0=0$ is shown in Fig.~\ref{fig50}.

%
%

\begin{figure}

a) \\
\centerline{
\psfig{file=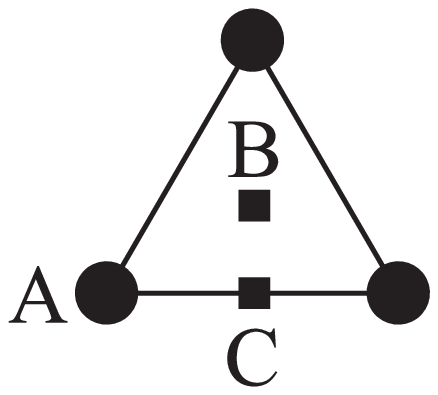,height=0.6in,bbllx=322pt,bblly=338pt,bburx=444pt,bbury=451pt}
}

b) \\
\centerline{
\psfig{file=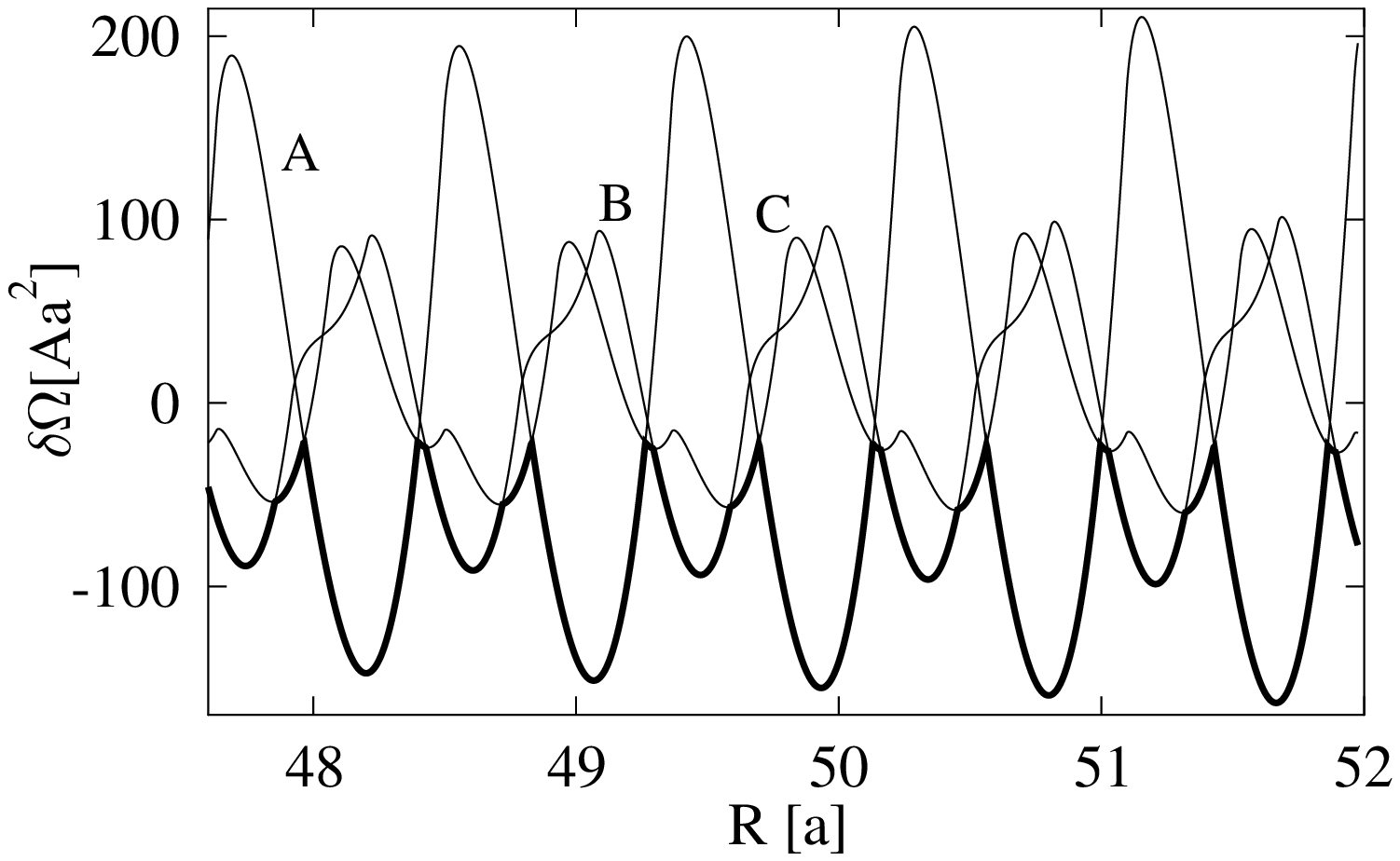,height=1.9in,bbllx=60pt,bblly=103pt,bburx=489pt,bbury=369pt}
}

\vspace{0.1in} 
\setlength{\columnwidth}{3.2in}
\centerline{\caption{
(a) Three point A, B, and C where the function $\overline{\Omega}_{\bbox{r}_0}$
has to have extrema due to symmetry.
(b) The oscillations of the thermodynamic potential 
for these three positions of the center. The minimum of these functions,
shown by the bold line, represents the actual value of the total energy
of the island. 
\label{fig60}
}}
\vspace{-0.1in}
\end{figure}

The next step is to minimize it with respect to $\bbox{r}_0$.
To do this we notice that $\overline{\delta \Omega _{\bbox{r}_0}}$
as a function of $\bbox{r}_0$ must have the same symmetry as the lattice.
Hence its extrema have to be located at points denoted in Fig.~\ref{fig60}a
by A, B, and C, unaffected by the transformations of the symmetry. 
Those are the points coinciding with a lattice site,
the center of the triangular face, 
and the point half-way between two lattice sites. 
Thus we actually need to make a choice between these three positions
of the center. This choice has to be made to minimize the energy 
$\overline{\delta \Omega _{\bbox{r}_0}}$. In Fig.~\ref{fig60}b this
minimization is done graphically. From the three curves 
$\overline{\delta \Omega _A}$, $\overline{\delta \Omega _B}$, 
and $\overline{\delta \Omega _C}$ for every value of $R$ the lowest one
is chosen. The resulting curve is shown by a bold line.
It indeed looks like the Kremlin wall structure obtained in
the numerical experiment.

We would like now to discuss the points of switching between two curves.
At these points the center of the island hops from one of the locations 
A, B, or C to another.
If the chemical potential is below or above one of these points the newly
added electrons are spread uniformly over the circumference of the island,
so that the center of mass does not move. But exactly at this point
$\sim N^{1/2}$ electrons travel from one side of the island to another shifting the
center of mass relative to the crystal. 
In addition to that at the point of transition $\sim N^{1/4}$ new
electrons enter the island. To understand this we recall that $N=-d\Omega /d\mu $.
As it is seen from Fig.~\ref{fig60}b at the point of transition
the first derivative of $\Omega$ has a discontinuity of the order of
\begin{equation}
\Delta\left[ -\frac{d\overline{\delta \Omega}} {d \mu} \right]
= - \frac {1}{2AR} \Delta \left[ \frac {d\overline{\delta \Omega}}{dR} \right] 
\sim \frac {Aa^2 N^{3/4}} {AaN^{1/2}\cdot a}
\sim N^{1/4}.
\end{equation} 
This implies that at this point a few peaks of the Coulomb 
blockade merge so that $\sim N^{1/4}$ electrons enter the island simultaneously.
The origin of this effect is in the electron attraction mediated by
the multi-polaronic effect associated with the confinement.
A simple toy model of this effect was suggested 
in Ref.~\onlinecite{Us_Phyl_Mag}.

%
%

The dependence of the total energy on the number of particles can
be obtained from our result using the small increment theorem,
stating that the small corrections to all the thermodynamic potentials are
the same:\cite{LL5,RemarkOnSmallIncrements}
\begin{equation}
\delta E (N) = \delta \Omega \left[ \mu (N) \right].
\end{equation}
This explains why in the dependence of the total energy on the number of
particles we also observe a Kremlin wall structure similar to Fig.~\ref{fig60}.
The amplitude and phase of the oscillations in Fig.~\ref{fig20} are in a very good
agreement with those given by 
Eq.~ (\ref{Delta_Omega}) minimized in the way shown in Fig.~\ref{fig60}.

We finally would like to notice that the appearance of a new terrace 
actually means starting a new crystalline row. This can be thought of as 
appearance of the pair of opposite dislocations on the boundary of the island.
Thus the variations of the total energy discussed above are associated 
with the periodic formation of the defects in the island. This line 
of thinking will be further developed for the case of the compressible crystal,
where these defects are situated inside the island.

%
%
%
%

\section{Coulomb Island: a Numerical Study}
\label{Coulomb}

\begin{figure}

a) \\
\centerline{
\psfig{file=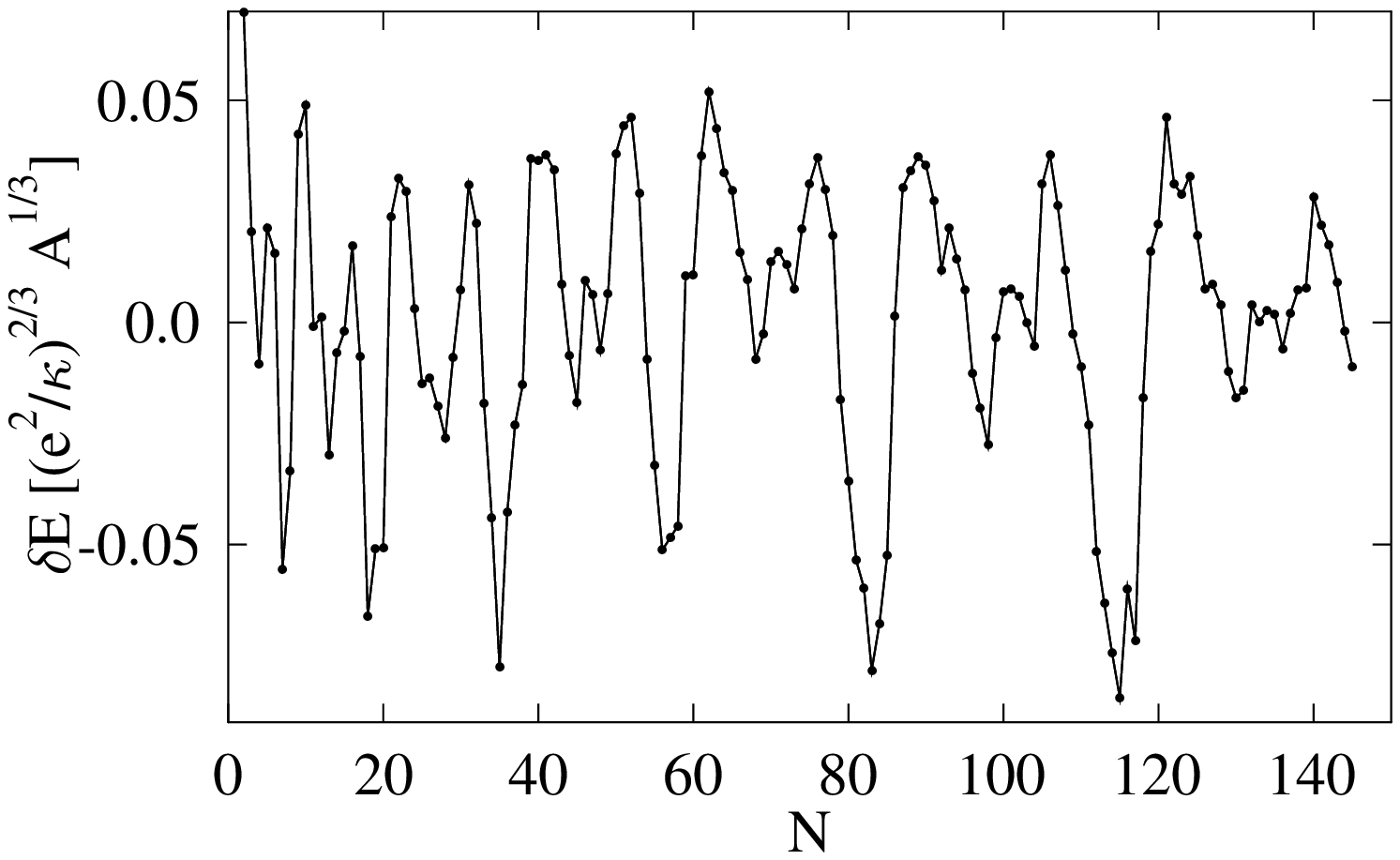,height=1.9in,bbllx=50pt,bblly=104pt,bburx=490pt,bbury=369pt}
}

b) \\
\centerline{
\psfig{file=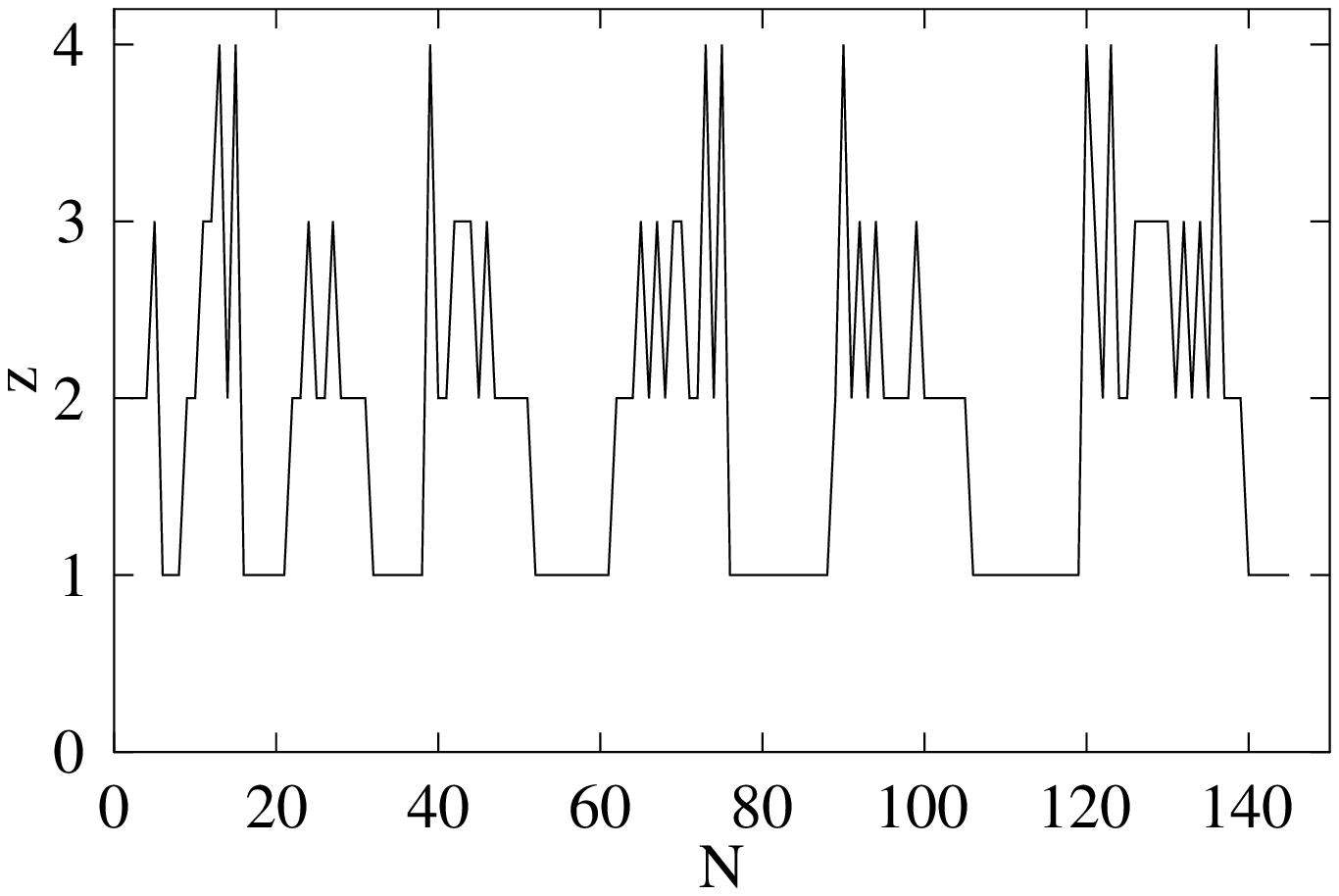,height=1.9in,bbllx=50pt,bblly=104pt,bburx=490pt,bbury=369pt}
}

\vspace{0.1in} 
\setlength{\columnwidth}{3.2in}
\centerline{\caption{
(a) Fluctuating part of energy of the Wigner crystal
island. (b) The number of particles adjacent to the center of the confinement. 
\label{fig65}
}}
\vspace{-0.1in}
\end{figure}

In this section we present the results of numerical solution
of problem (\ref{Energy}) with the unscreened Coulomb interaction 
(\ref{Coulomb_int}). To obtain these results we employed the 
numerical technique described in the previous section.
The total energy $E(N)$ resulting from such a calculation 
was split into the smooth and the fluctuating components in the
way described in the previous section. The smooth component was chosen to
have the form\cite{Us_Phyl_Mag}
\begin{equation}
\begin{array}{ll}
{\displaystyle
\bar{E}\left( N \right) = } & 
{ \left( e^2/\kappa \right) ^{2/3} A^{1/3} \left( \displaystyle \eta_1 N^{5/3}
+ \right.} \\ 
 {~} & {\left. \displaystyle + 
\eta_2 N^{7/6} + \eta_3 N^{2/3} + 
\eta_4 N^{7/15} + \ldots \right)} ,
\end{array}
\label{E_ave}
\end{equation}
where $\eta_i$ are some constants.
The first term in this series is the electrostatic energy.
The next three terms are
the correlation energy, the overscreening energy associated 
with the screening of the external potential by the Wigner crystal, 
and the surface energy. The coefficients $\eta_i$ can be
found from the best fit to the numerical data:
$\eta_1 = 6/5\cdot \left( {3\pi}/8 \right)^{2/3}$ , 
$\eta_2 = -1.0992$ , $\eta_3 = -0.3520$ , $\eta_4 = 0.1499$.
The fluctuating part is displayed in Fig.~\ref{fig65}a.
%
%

The curve in Fig.~\ref{fig65}a has a quasi-periodic structure
similar to the incompressible case (see Fig.~\ref{fig20}).
It consists of the sequence of interchanging deep and shallow minima. 
The positions of minima on this graph almost
exactly coincide with those for the case of incompressible island.
The amplitude of the oscillations does not change with $N$ appreciably
and is $\sim 0.1e^2/\kappa a$. The charging energy $\Delta (N)$
experiences fluctuations $\approx 15\%$ correlated with
the positions of the maxima and minima.

The similarity of Figures~\ref{fig20} and \ref{fig65}a calls for the
conclusion that the fluctuations in both cases are of the same origin.
To test this idea we studied such a quantity as the number of 
electrons adjacent to the center of the confinement.   
Those are defined as electrons nearest to the center,
with dispersion in the distance to the center being
less than $1/3$ of the lattice spacing.
They represent the first crystalline shell closest to the center.
The number of these particles is shown in Fig.~\ref{fig65}b as 
a function of the total number of electrons in the island.
The first thing to notice is the correlation of the latter graph
with the fluctuations of energy in Fig.~\ref{fig65}a. The deep minima
in the energy are associated with having one electron next to the center.
The shallow minima mostly correspond to having three electrons
at the center. This in fact means that the center of the 
confinement is situated in the center of the triangular crystalline face.
Maxima of energy usually correspond to having two or four electrons
next to the center. This is equivalent to saying that the center is in the
middle of a crystalline bond. Thus we see that this correspondence
is exactly the same as switching between terms A, B, and C in the incompressible
case (see Fig.~\ref{fig60}). This analogy will be further
developed in Section~\ref{Compr_Isl}. 

\begin{figure}
\centerline{
\psfig{file=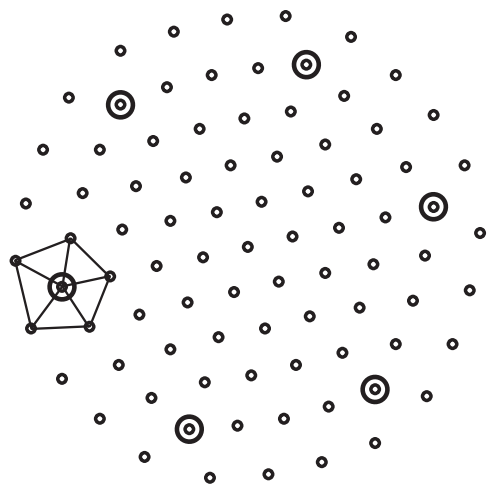,height=2.5in,bbllx=193pt,bblly=247pt,bburx=402pt,bbury=454pt}
}
\vspace{0.1in} 
\setlength{\columnwidth}{3.2in}
\centerline{\caption{
The magic number configuration \protect{$N=85$}.
Six \protect{five}-coordinated particles associated with
disclinations are marked by rings.
Triangulation of the neighborhood of one of them is shown 
explicitly.
\label{fig70}
}}
\vspace{-0.1in}
\end{figure}

\begin{figure}
\centerline{
\psfig{file=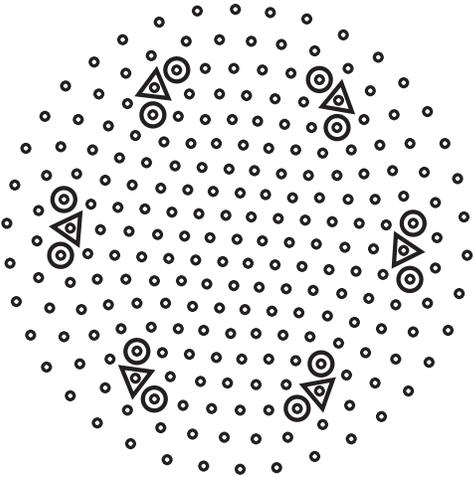,height=2.5in,bbllx=160pt,bblly=371pt,bburx=382pt,bbury=596pt}
}
\vspace{0.1in} 
\setlength{\columnwidth}{3.2in}
\centerline{\caption{
The electron configuration for \protect{$N=235$}.
The particles having five and seven nearest neighbors are shown by the rings and
the triangles respectively.
\label{fig72}
}}
\vspace{-0.1in}
\end{figure}

Let us now examine the conformations of electrons.
Some of them are shown in Figures~\ref{fig70} and \ref{fig72}.
The first thing obvious from these Figures is that the surface of
the island is not so rough as it was in the incompressible case
(compare to Fig.~\ref{fig10}). At large $N$ the surface contains no defects. 
The defects reside in the interior instead. 

%
%

Let us define these defects.
Elementary defect in a 2D lattice is disclination. 
The only possible form of such a defect in 
2D crystals is the so-called wedge disclination. 
It can be viewed as a wedge which is removed (inserted) from
the crystal. The ``charge'' of the disclination is the 
angle, formed by the wedge. The minimum possible
disclination charge for a triangular lattice is $\pi/3$. 
The disclinations can be identified with the particles having an
anomalous coordination number. In the triangular crystal
the cores of the positive or negative disclinations are 
associated with the particles having five or seven nearest neighbors respectively,
the normal coordination being six.
Some examples of such defects are shown in Fig.~\ref{fig70}. 

Dislocations are the pairs of positive and negative disclinations 
forming a dipole. They can be seen in Fig.~\ref{fig72}.
When number of electrons in the island is less than some 
critical value $\tilde{N} \approx 150$,
there exist highly symmetric electron configurations free of
dislocations. These configurations can be realized only
at some distinct values of $N = N_m = 7, 19, 35, 55, 85$, which we call 
{\em magic numbers}.
One of such magic number configurations $N_m=85$ is demonstrated
in Fig.~\ref{fig70}. On the other hand if $N > \tilde{N}$ the dislocations are always
present and the magic number configurations never exist.

%
%

We would like to stress again that all these defects are not
observed directly on the surface. 
Moreover, if the number of electrons in the
island is large, they do not come close to the surface. 
The same is true about the central
region of the island. It is also usually free of defects. All the
irregularities in the lattice are normally observed in a ring of width
$(2\div 3)a$ at a fixed distance from the edge. 

We would like to discuss now the accuracy of the results we have obtained.
We believe that the total energy for the cases shown in Fig.~\ref{fig65}a
is calculated with precision $10^{-6} (e^2/\kappa)^{2/3} A^{11/3}$. 
This was tested by reruns 
(starting from different initial conditions)
and comparison with the results of
simulated annealing. The configurations however cannot be reproduced
reliably. The thing is that several completely different
configurations for the same number of electron can have very close energies,
within $10^{-6} (e^2/\kappa)^{2/3} A^{11/3}$ from each other.
However the general features of all these low energy configurations 
described above remain true. Thus, although we cannot reproduce the
position of every individual defect reliably, we believe
that we can say that in general defects reside in a ring concentric
to the surface of the width of a few lattice constants. 

We will see in Section~\ref{Compr_Isl} that the interaction of defects is
crucial in understanding of the energy fluctuations. 
In the next section we discuss the general properties
of the defect distributions. These properties are essential
for the arguments given in Section~\ref{Compr_Isl}.

%
%
%
%

\section{Lattice Defects in the Circular Wigner Crystal Island}
\label{Defects}

When the triangular crystal is packed into the island of circular form 
the obvious incompatibility of these two structures
leads to the appearance of
the lattice defects: disclinations and dislocations.
The important difference between the former and the
latter is that disclinations are {\em always} present
in the ground state. The number of disclinations 
in the island is determined by the Euler's theorem
(see below) and, hence, cannot be changed.
Dislocations on the other hand may or may not
appear depending on whether they are 
energetically favorable. 
Let us first explain why the disclinations have to appear
in the ground state.

The disclinations within some region of lattice
can be identified by the number of $\pi /3$ turns that
one has to make to walk around it. Normally one has to
make minimum six such turns (consider for example the central
point in Fig.~\ref{fig70}). For the region having $N_c$ such defects
inside, the minimum number of $\pi /3$ turns is $6-N_c$.
Look e.g. at the pentagon in Fig.~\ref{fig70}. It obviously
contains one disclination, because the number of the aforementioned turns
is five. Let us now mentally walk around the {\em whole} circular sample
along its edge.
As the surface of the island is circular we do not make
any turns {\em at all}. Hence the total number of $\pi /3$ disclinations 
is exactly six:
\begin{equation}
N_c = 6.
\label{Nd}
\end{equation}
In the simplest possible case
shown in Fig.~\ref{fig70} there are only
six disclinations in the sample.
If both positive (removal of $\pi /3$ wedge) and negative
(insertion of such a wedge) disclinations are present, 
the total disclination charge should be calculated accounting
for the sign of these defects:
\begin{equation}
N_c \equiv N_c^{+} - N_c^{-},
\label{Nd_def}
\end{equation}
where $N_c^{+}$ and $N_c^{-}$ are 
the numbers of positive and negative
defects correspondingly.
Such complex cases are realized for example when there are a few dislocations
in the island. Dislocation is a pair of positive and negative
disclinations bound together to form a dipole. Hence
addition of a dislocation to the sample increases both 
$N_c^{+}$ and $N_c^{-}$ keeping their
difference the same.
Consequently the number of dislocations is not
controlled by the topological constraints expressed 
by Eq.~(\ref{Nd}).
This equation can also be proved using Euler's theorem.
The proof can be found in Appendix ~\ref{App_B}.


Now let us turn to the dislocations. At zero temperature they
embody inelastic deformations.
There are two main reasons for existence of such deformations:
inhomogeneity of the concentration of electrons
and the presence of disclinations.
The first reason is not universal: the exact profile of density
is determined by the form of confinement and interaction potential.
The second one is universal, because the presence of disclinations
is required by theorem~(\ref{Nd}).

The dislocations produced by the varying crystal density
were recently discussed by Nazarov.~\cite{Nazarov96}
He noticed that the dislocation density must be equal to the 
gradient of the reciprocal lattice constant:
\begin{equation}
\bbox{b}\left( \bbox{r} \right) = a\left( \bbox{r} \right) 
\bbox{\hat{z}} \times \bbox{\nabla} a^{-1} \left( \bbox{r} \right) .
\label{b_r}
\end{equation}
Here $\bbox{b}\left( \bbox{r} \right)$ is the density 
of Burgers vector, $\bbox{\hat{z}}$ is the 
unit vector perpendicular to the surface, and
$a = \sqrt{2/n\left( \bbox{r} \right) \sqrt{3}}$
is the varying in space lattice constant. 
This formula is easy to understand.
First we notice that a dislocation adds an extra row
to the lattice. Hence the density of dislocations 
is given by the rate of change of concentration of these
planes equal to $1/a$. To obtain the Burgers
vector density we multiply this gradient by the 
elementary Burgers vector $a$ (see also Ref.~\onlinecite{RemarkOnNazarov}).

This formula has an interesting implication for the question
of applicability of the elasticity theory to
the Wigner crystal. Consider a sample of size $L$.
Assume that there is a variation of concentration in the
sample induced by an external source $\delta n \sim u_{ik} n$,
where $u_{ik}$ is the strain tensor.
Let us find the value of $u_{ik}$ at which dislocations start to appear.
This would imply that inelastic (plastic) deformations 
occur and the elasticity theory brakes down.
The energy of purely elastic deformations is
\begin{equation}
E_{\rm el}\sim Y L^2 u_{ik}^2.
\label{Elastic_Energy}
\end{equation}
Here $Y$ is the Young's modulus.
For inelastic deformations we have
\begin{equation}
E_{\rm inel} = E_c N_d,
\label{Inelastic_Energy}
\end{equation}
where $E_c$ is the dislocation core energy and $N_d$ is the total number of
dislocations in the sample. Eq.~(\ref{b_r}) provides the following estimate for
the latter: 
\begin{equation}
N_d \sim L^2 \frac{\delta n }{ nLa } \sim u_{ik} \frac La.
\end{equation}
Comparing the elastic and inelastic energies we conclude
that the deformations are completely elastic when 
\begin{equation}
u_{ik} \lesssim \frac {E_c}{YaL}.
\label{u_ik}
\end{equation}
For the Wigner crystal with the Coulomb interaction using
$E_c \sim e^2/\kappa a$, $Y \sim e^2/\kappa a^3$, and $u_{ik}\sim u/L$,
where $u$ is the characteristic displacement, we obtain
\begin{equation}
u \lesssim a.
\label{The_condition}
\end{equation}
Hence equilibrium elastic displacements cannot exceed the lattice spacing
for the Wigner crystal.
This agrees with the known result that the elasticity theory has 
a zero radius of convergence with respect to the strain tensor.~\cite{Sethna96}
In effect we conclude that the radius of convergence depends
on the spacial scale of the problem and in the macroscopic limit
$L\rightarrow \infty$ is indeed zero, according to (\ref{u_ik}).
This surprising result emphasizes the difference between
electron crystal and crystal consisting of heavy particles (atoms). 
In an electron crystal the 
relaxation time provided by tunneling is smaller than the time of experiment and, 
hence, the system can find the ground state.
In an atomic crystal the relaxation time is typically much larger than
the experimental time and the system can remain in the metastable 
state described by the elasticity theory.  

Below we compare the number of dislocations of two different origin.
First we calculate the total number of dislocations 
produced by the non-uniformity of the electron density.
We consider the case of the Coulomb interaction.
For this case the corresponding electrostatic problem (ignoring the discreteness 
of the charge of electrons) can be solved exactly.~\cite{Sneddon66}
The solution for the density of electrons can be shown to be a ``hemisphere'':
\begin{equation}
\begin{array}{l}
{\displaystyle
n\left( r\right) = n_0\sqrt{1- \frac {r^2}{R^2}}} \\ \\
{\displaystyle
n_0 = \frac {4A\kappa R}{\pi^2e^2} ,~
R = \left( \frac {3\pi Ne^2}{8A\kappa} \right)^{1/3}.
}
\end{array}
\label{n_r}
\end{equation}
According to Eq.~(\ref{b_r}) the density of dislocations induced by varying 
electron density is
\begin{equation}
n_d(r) = \frac {b_{\phi}(r)}{h} = \sqrt{ \frac {2 n_0} {\sqrt{3}} }
\frac {r} {R^2\left(1-r^2/R^2\right)^{3/4} }
\label{n_d}
\end{equation}
Here $h=a\sqrt{3}/2$ is the elementary Burgers vector, equal to 
the distance between crystalline rows.
The total number of dislocations is readily obtained by integrating this distribution:
\begin{equation}
N_{d1} = \int_0^R 2\pi rdr n_d(r) \approx 4.08 \cdot N^{1/2}. 
\label{N_d1_1}
\end{equation}

The electrostatic formula (\ref{n_r}) is correct if the number of electrons is large. 
In the opposite case of a small island
the density profile is very far from being a hemisphere. 
In this case the density is almost constant.\cite{Us_Phyl_Mag} 
This is obvious from looking at Figs.~\ref{fig70} and \ref{fig72} above.
To evaluate the variation of the electron density 
we will find the lattice constant on the surface
of the island and compare it to that in the center.
To this end we notice that
near the edge the density of dislocations given by Eq.~(\ref{n_d})
would become infinitely large. However it cannot exceed the density
of electron themselves, given by Eq.~(\ref{n_r}). The position
of the edge of the crystal can therefore be determined matching
the dislocation density and the density of electrons.
As a result we obtain for the lattice constant on the edge
the following expression:
\begin{equation}
\lambda \sim a_0^{4/5}R^{1/5} \propto N^{-1/15}.
\label{lambda_par}
\end{equation}
This expression agrees with conclusions of Ref.~\onlinecite{Us_Phyl_Mag}
obtained in a different way.
Our numerical data agree very well with this theory and give the following 
coefficient:
\begin{equation}
\begin{array}{c}
{\lambda = 0.88 a_0^{4/5}R^{1/5},} 
\end{array}
\label{lambda}
\end{equation}
The interelectron distance on the edge given by this formula
differs very slightly from the lattice constant in the
center of the sample if the radius of the island is not too large.
To evaluate the number of dislocations for a small island
we approximate the electron density profile with a parabola:
\begin{equation}
n(r) \simeq n_0\left[ 1-\frac {r^2}{R^2} \left( 1-\frac {a_0^2} {\lambda^2}
\right) \right].
\end{equation} 
Applying Eq.~(\ref{b_r}) to this expression we obtain 
\begin{equation}
N_{d1} \approx 2.05 \cdot N^{1/2} \left( 1-\frac {a_0^2} {\lambda^2} \right)
\label{N_d1_2}
\end{equation}
In the consideration below we will assume that the number of electrons in the 
island is not too large. Therefore the electron density is almost uniform.

As it was mentioned above, another reason for appearance of dislocations
is the stress, produced by disclinations. The latter
deform the lattice enormously. 
These deformations
can be significantly reduced by introducing dislocations into the lattice.
This process is usually referred to as screening.
The screening of disclinations by dislocations
has been studied before in the context of the hexatic liquid --
homogeneous liquid transition.\cite{Lubensky95} The screening is that case
is accomplished by polarization of the thermally excited dislocations
above the Kosterlitz-Thouless transition.
It was therefore treated in the framework of the linear Debye-Huckel approximation.
Here we deal with the case when there are {\em no} thermally excited
dislocations and all the Burgers vector density necessary for 
screening exists due to the appearance of new dislocations.
 
The phenomenon can be most easily understood considering 
the total disclination density:
\cite{Lubensky95}
\begin{equation}
s_{\rm tot}(\bbox{r}) = s(\bbox{r}) - \varepsilon_{ik} \nabla_i b_k(\bbox{r}),
\label{Generalized}
\end{equation}
where the first term is the density of free disclinations 
while the second is the density of disclinations induced
by the varying in space density of dislocations (the latter are the dipoles
formed from the former). The elastic energy of the crystal can
be written in terms of this total defect density:
\begin{equation}
E = \frac Y2 \int \frac {d^2q}{(2\pi )^2} 
\frac {\left| s_{\rm tot}(\bbox{q})\right|^2} {q^4},
\end{equation}
where $Y$ is the Young's modulus. For case of the Coulomb crystal~\cite{Fisher79}
$Y=\alpha e^2n^{3/2} /\kappa$, where $\alpha=0.9804$. We notice
that for the large distances $q \rightarrow 0$ the numerator in this
expression is pushed to zero by the $q^4$ term contained in the denominator.
Thus, ignoring the effects at the distances $r \sim a$, 
one can write the condition of the perfect screening:
\begin{equation}
s_{\rm tot}(\bbox{r}) = 0.
\label{Complete_scr}
\end{equation}
It is instructive now to consider one disclination in the 
center of an infinite sample: $s(\bbox{r})=s_0\delta(\bbox{r})$.
Equations (\ref{Complete_scr}) and (\ref{Generalized}) have in general many
solutions. Two of them are easy to guess: 
$b_{1\phi} = s_0/2\pi r$, $b_{1r}=0$ and $b_{2x}=0$, $b_{2y}=s_0 \Theta(x) \delta(y)$. 
They represent the Burgers vector rotating around the disclination and the
solution in the form of a grain boundary respectively. 
They are different by the longitudinal
Burgers vector density: $\bbox{b_1} = \bbox{b_2} + \nabla f$, where $f$ is
some scalar function, and hence have the same elastic energy. This 
non-uniqueness is
a consequence of essentially mean-field character of Eq. (\ref{Complete_scr}).
Discreteness of the dislocations removes this degeneracy in
favor of the grain boundary. The idea is that
the fluctuations of the local elastic energy caused by the discreteness
of the defects can be estimated as $E\sim Ya^2\ln (\left< r\right>/ a)$
per defect,
where $\left< r\right>$ is the average distance between them. 
In the former distribution $\bbox{b_1}$ 
this average distance grows with the size of the
sample $\left< r\right> \sim \sqrt{L a}$. In the case of the grain boundary
the average distance is maintained of the order of the lattice spacing.
Hence the self-energy of defects in the latter case does not grow with 
the size of the sample. 

\begin{figure}
\centerline{
\psfig{file=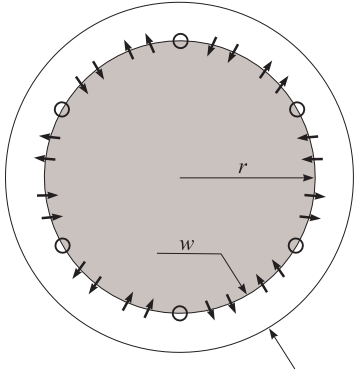,height=1.9in,bbllx=120pt,bblly=234pt,bburx=498pt,bbury=625pt}
}
\vspace{0.1in} 
\setlength{\columnwidth}{3.2in}
\centerline{\caption{
The distribution of defects in a compressible island. 
Rings and arrows show the positions of disclinations and the direction of the 
Burgers vector density, respectively. The gray disk shows the 
free of defects central region.
\label{fig80}
}}
\vspace{-0.1in}
\end{figure}

The next logical step is to consider six disclinations in the circular
crystalline sample. 
A possible solution for the distribution 
of defects is shown in Fig.~\ref{fig80}. It is prompted by the results of
the numerical simulations showing that in the majority of cases the defects
(dislocations and disclinations) are situated in a ring concentric
to the surface of the island. The regions adjacent to the center and to 
the edge of the island appear to be free of defects. 
We assume that the disclinations form a figure close to a perfect hexagon.
The dislocations form grain boundaries connecting the disclinations.
This is done to smear out the charge of disclinations in accordance with the
screening theory expressed by Eq.~(\ref{Complete_scr}).  

%
%

Next we calculate the distance from the surface to the layer
of defects. 
We consider the case of a uniform crystal.
The central region of the island can be formed with no deformations in it.
It can be thought of as a piece of the uniform crystal
having a circular form, the same as the one considered in 
Sections~\ref{Inc_Isl} and \ref{Inc_Isl_ATheory}.
The central region is shown in Fig.~\ref{fig80} by a gray disk.
The region between the surface of the island and the layer of defects
is free of defects too.
It cannot be free of elastic deformations however.
These deformations are the same as those of a 2D rectangular elastic rod,
two opposite edges of which are glued together.
To find the optimum thickness of this rod $w$ (see Fig.~\ref{fig80})
one has to balance the energy associated with this bending, which tends 
to reduce $w$, and the surface energy of the grain boundary. The latter
is proportional to the length of the grain boundary and therefore has a tendency to
increase $w$.
The bending energy can be calculated from the elasticity theory~\cite{LL7_2}
\begin{equation}
E_{\rm el} = \frac {\pi}{12} Y \frac {w^3}R.
\label{E_el_ring}
\end{equation}
The surface energy of the grain boundary can be estimated from the core
energies of the dislocations:
\begin{equation}
E_{\rm gb} \approx E_c N_d
\label{E_grain_b}
\end{equation} 
The total number of dislocation follows from elementary geometric
calculation done for the incompressible crystalline region of radius $r=R-w$:
\begin{equation}
N_{d2} = 12(1-\sqrt{3}/2) r/h,
\label{N_d_total}
\end{equation}
where $h=a\sqrt{3}/2$. Minimizing the total energy $E=E_{\rm el}+E_{\rm gb}$
we find the optimum at:
\begin{equation}
w \approx 1.5 \sqrt{\frac {E_cR} {Ya} } \approx 0.5 \sqrt{Ra}.
\label{w_ring}
\end{equation} 
In the derivations of (\ref{w_ring}) we used $E_c=0.11e^2n^{1/2}/\kappa$ 
and $Y=0.9804e^2n^{3/2}/\kappa$ (Ref.~\onlinecite{Fisher79}).
This result can be also obtained from our condition of stability of a crystal
(\ref{The_condition}). Indeed the strain tensor in the ring can be estimated
as $u_{\phi \phi} \sim w/R$. The corresponding displacement $u \sim w^2/R$ cannot
exceed the lattice spacing $a$, according to Eq.~(\ref{The_condition}).
The estimate for the width of the ring $w$ obtained from this argument is
consistent with Eq.~(\ref{w_ring}).

Now we would like to find the condition 
at which the dislocations stay on the surface of the island.
We will consider the case of the short-range interaction (\ref{Exp_int}).
To this end one has to compare the energy of the defects on the boundary
with that in the bulk of the crystal. The former is associated with the
roughness of the surface (see Fig.~\ref{fig10}). It can be estimated 
as a product of the total number of particles there $\sqrt{N}$, the typical
force acting on a particle on the surface $F(R) = 2AR$, and the characteristic 
deviation of the shape of the surface from the circle, given by the lattice constant:
$\delta E_{\rm sur} \sim \sqrt{N} \cdot F(R) \cdot a$.
The energy of the defects inside the island
is given by $\delta E_{\rm bulk} \sim  \sqrt{N} \cdot Ya^2 \ln N$.
The first term in the right-hand side is the number of defects 
and the second one is the typical energy per dislocation.
It is estimated as the typical interaction energy of two dislocations. 
Note that we neglect the core
energy of dislocations as it is small compared to the 
$Ya^2$ for the crystal with short-range interaction.
Indeed the former is of the order of the correlation energy:
$E_c \sim U_0\exp (-a/s)$.
The Young's modulus for the considered system can be estimated as the
second derivative of the interaction potential between two particles:
\begin{equation}
Y \sim \frac {U_0}{s^2} \exp \left( -\frac as \right).
\label{Young}
\end{equation}
It is clear then that $E_c/Ya^2 \sim s^2/a^2 \ll 1$.
Comparison of $\delta E_{\rm sur}$ with $\delta E_{\rm bulk}$ 
gives the condition that the defects stay on the surface:
\begin{equation}
Y\ln N \gg A \sqrt{N}.
\label{Cond1}
\end{equation}     
The coefficient $A$ can be conveniently related to the 
interaction strength by balancing the 
forces acting on a particle on the surface: $AR \sim U_0/s\cdot \exp(-a/s)$.
Combining this equation with Eq.~(\ref{Young}) and (\ref{Cond1}) 
we obtain the following condition for the defects staying on the boundary:
\begin{equation}
a \ln N \gg s
\end{equation}
This coincides with the condition of uniformity of the crystal 
(see Section~\ref{Inc_Isl_ATheory}). Note that this entire consideration 
is based on the assumption of uniformity of the crystal. 

Finally in this section we evaluate the critical number of electrons $N^*$ 
at which the number of dislocations due to the 
inhomogeneity of the electron density $N_{d1}$ and produced by the screening 
of disclinations $N_{d2}$ become equal. 
To do so we compare Eq.~(\ref{N_d1_2}) with Eq.~(\ref{N_d_total}).
They become equal if $\lambda \approx 1.55 a_0$ or, using Eq.~(\ref{lambda_par})
\begin{equation}
N=N^*\simeq 700.
\end{equation}
Therefore if the number of electrons in the island is smaller that $N^*$ we
dislocations are mostly due to the screening of disclinations and are arranged into
the grain boundary (see Fig.~\ref{fig80}). In the opposite
case the dislocations are generated in the interior 
according to Eq.~(\ref{b_r}).

The above consideration is essentially mean-field, treating the 
density of dislocations  as a continuous quantity. In the next
section we give an argument that the  discreteness
of dislocations is responsible for the fluctuations of the elastic
energy of the Wigner lattice.

%
%
%
%

\section{Elastic Blockade}
\label{Compr_Isl}

The number of dislocations in the Wigner crystal island is of the
order of the number of crystalline rows in it [see Eq.~(\ref{N_d_total})]:
\begin{equation}
N_d \sim R/a \sim \sqrt{N}.
\end{equation}
Hence, it grows while the island in filled with electrons. On
the average one dislocation appears as $N$ increases by $\delta N\sim \sqrt{N}$. 
The elastic energy stored due to the
deviation of the number of dislocations from the average is
relaxed when a new one is added. This phenomenon is similar
to the Coulomb blockade, with the words ``electron'' and ``electrostatic energy''
replaced by ``dislocation'' and ``elastic energy''. 

Using this analogy to the Coulomb blockade one can easily estimate
the order of the fluctuations of energy. In the Coulomb blockade case
the fluctuations of the electrostatic energy are given by the following expression:
\begin{equation}
\delta E = \frac {e^2}{2C} \delta N^2,
\label{Coulomb_blockade}
\end{equation}
where $- 1/2 \le \delta N < 1/2$ is the deviation of the number of electrons
in the dot from the average, determined by the gate voltage. 
Following the convention of the above mapping one has to replace
$e^2/C$ by the characteristic energy of interaction of two dislocations
in the island $U_0 \sim Ya^2 \sim e^2/\kappa a$.
One also has to replace $\delta N$ by the deviation of the number of
dislocations from the average $\delta N_d \sim \delta N / \sqrt{N}$.
As a result we obtain:
\begin{equation}
\delta E_{\rm el} = \alpha \frac {e^2}{\kappa a} \frac {\delta N^2}{N}.
\label{Main_Result}
\end{equation}
Here $\alpha\approx 0.5$ is a numerical constant.
The approximate value of this constant is obtained from 
the numerical results of Section~\ref{Coulomb}.
As the total period of such an ``elastic'' blockade is $T_N \sim \sqrt N$,
the maximum fluctuation of energy that can be reached is 
$\delta E_{\rm max}\sim \alpha e^2/\kappa a$. 

Another implication of this analogy is the appearance of the
elastic blockade peaks. 
In the case of Coulomb blockade the intersection of two terms 
described by Eq.~(\ref{Coulomb_blockade}) for different numbers of electrons in the 
dot produce a spike in the conductance through the dot.
At this point a new electron enters the dot.
For the ``elastic'' blockade at the similar point a new dislocation enters the 
crystalline island. The chemical potential of an electron has
a discontinuity of the order of 
$\Delta \mu \sim dE/dN \sim - \alpha e^2/\kappa a T_N \sim - \alpha e^2/R$.
The charging energy at this point fluctuates by a value
\begin{equation}
\delta \Delta = \frac {d\delta \mu}{dN} = \frac {\Delta \mu}1 \sim 
- \alpha \frac {e^2}{\kappa R},
\label{delta_Delta}
\end{equation}
which is of the order of the average $\overline{\Delta} = \beta e^2/\kappa R$,
$\beta$ being another constant. 
This however does not lead to the bunching in the charging spectrum
as it did in the case of the short-range interaction, because of the smallness 
of the numerical constant $\alpha$.

Let us consider the elastic blockade in more detail.
In the previous section we derived the condition at which the
crystal can be considered to be uniform. If the number of electrons
in the island is less than some critical value $N^* \simeq 700$, the
variations in the density of electrons can be ignored and the number of
dislocations associated with the variable density is small.
Below we examine these two regimes separately.

{\em i) Almost uniform crystal: $N \lesssim N^{*}$.}

As it follows from the previous section in this regime the 
density of electrons in the island is almost uniform.
The crystal is packed into the circular form by forming two
monocrystals: the almost circular internal region, free of elastic
deformations, similar to the incompressible island; and the
external ring (see Fig.~\ref{fig80}). The interface between these two
monocrystals is a grain boundary, which can be viewed as a string of dislocations.

Assume now that the position of the center of the confinement 
relative to the lattice $\bbox{r}_0$ is fixed. This coordinate has been introduced
earlier in Section~\ref{Inc_Isl_ATheory}.  
One can then calculate the energy
of the island as a function of the number of electrons and the 
position of the center $E_{\bbox{r}_0}(N)$. 
Due to the symmetry considerations given in Section~\ref{Inc_Isl_ATheory} this
function has extrema at the point
A, B, and C shown in Fig.~\ref{fig60}a
similar to the incompressible island. Filling of the island 
results in switching among three energetic branches $E_{A}(N)$, $E_{B}(N)$,
and $E_{C}(N)$, every time choosing the lowest.
Below in this subsection we estimate 
the correction to the total energy $\delta E_{\bbox{r}_0}(N)$.

The grain boundary has a tendency 
to be at a fixed distance from the center $r=R-w$, calculated in 
Section~\ref{Defects}. If position of the center relative to the
crystal is fixed, filling of the island brings about expanding
of the mean grain boundary position with respect to the crystal.
Hence, periodically the grain boundary has to intersect a new crystalline
row. At this moment in the corresponding incompressible problem
a new terrace appears on the boundary of the island. In the 
Coulomb problem the grain boundary is submerged into the island. 
Therefore at this moment in the compressible case a new row is added to the island.
Since the termination points of the new row can be thought of
as dislocations, we can say that a new pair of opposite dislocations
appears on the grain boundary (see Fig.~\ref{fig80} with the gray
internal region shown in more detail in Fig.~\ref{fig30}).

The corresponding fluctuation of energy can be evaluated as
the energy of interaction of two opposite dislocations
having charge $\delta N_d = N_d - \overline{N}_d$, where
$N_d$ is the actual number of dislocations in the neighborhood
of the new row, $\overline{N}_d$ is the average one. This
energy is given by:\cite{Lubensky95}
\begin{equation}
\delta E_0 \simeq \frac {Yh^2}{4\pi} \delta N_d^2 
\ln \left( \frac{ l }{h} \right).
\label{Delta_E_beg}
\end{equation}  
The typical distance between defects $l$ is given by Eq.~(\ref{terrace_length}):
$l\sim \sqrt{R\cdot a}$. The average number of dislocations is given by
the number of crystalline rows $\overline{N}_d = \sum _i (r-\bbox{\hat{e}}_i \bbox{r}_0)/h$, 
$\bbox{\hat{e}}_i$
being the unit vector normal to the series of rows considered. Therefore
\begin{equation}
\delta N_d = \sum _{i=1}^6 \left\{ \frac {r-\bbox{\hat{e}}_i \bbox{r}_0} {h} \right\} - \frac{1}{2},
\end{equation}
where by $\{\cdots \}$ we assume taking the fractional part.
The oscillations of energy are given by the formula similar to 
Eq.~(\ref{Delta_Omega})
\begin{equation}
\delta E_{\bbox{r}_0} = \sum_{i=1}^6
\delta E_0(r-\bbox{\hat{e}}_i \bbox{r}_0),
\label{Delta_E}
\end{equation}
where
\begin{equation}
\delta E_0(r)  \simeq \frac {Yh^2} {16 \pi} 
\left[\left( \left\{ \frac rh \right\} - \frac 12 \right)^2 - \frac 1{12} \right] 
\ln N 
\label{Delta_E_0}
\end{equation}
are the fluctuations of energy given by (\ref{Delta_E_beg}).
The last term in the square brackets is chosen so that
$\overline {\delta E_0 (r)} = 0$. 

\begin{figure}
\centerline{
\psfig{file=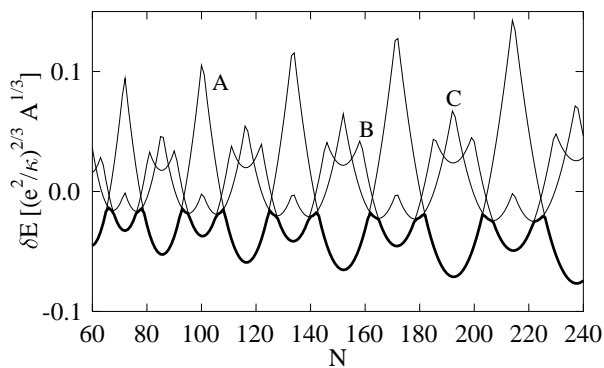,height=1.9in,bbllx=50pt,bblly=104pt,bburx=490pt,bbury=369pt}
}
\vspace{0.1in} 
\setlength{\columnwidth}{3.2in}
\centerline{\caption{
The fluctuating part of the energy of the Wigner crystal island
calculated from the interaction of dislocations. 
\label{fig90}
}}
\vspace{-0.1in}
\end{figure}

The oscillations of energy calculated in this way for three 
positions of the center $\bbox{r}_0 = $ A, B, and C are shown in Fig.~\ref{fig90}. 

%
%
As usually the lowest one is to be chosen. Comparison to the
numerical experiment (Fig.~\ref{fig65}) shows, that this
model predicts very well the phase and the shape of the oscillations.
The amplitude however is smaller by a factor of $2 \div 2.5$
than observed in the simulations. This difference can be 
explained by the influence of the boundary of the sample on
the interaction energy of dislocations (\ref{Delta_E_beg}).
Indeed the boundary of the island is at the distance 
$w \approx 0.5 \sqrt{R\cdot a}$ from the ring of dislocations. 
This scale is smaller than the characteristic distance between dislocations 
$l\approx (2\div 3)\sqrt{R\cdot a}$.
Hence the boundary of the island should play an important role in the
interaction of dislocations. 

To understand the character of the correction we notice
that the boundary in the numerical experiment is almost not deformed.
The island tends to preserve its circular shape.    
Hence we can assume that the normal displacement of the boundary is zero.
Assume now that the dislocations are much farther away from each other
than from the boundary $l \gg w$. In practice we have $l \approx 5 w$.
The problem of interaction of the pair of dislocations in the
presence of the boundary can then be solved using the method of images.
It is easy to see that at the zero normal displacement boundary condition
a dislocation with the Burgers vector perpendicular to the boundary 
produces an image of the same sign. Indeed 
such a dislocation is a crystalline semi-row, parallel to the surface. 
The zero normal displacement boundary condition
can be realized by putting a parallel semi-row on the other side
of the boundary. Hence effectively the boundary doubles the charge of each
dislocation. The interaction energy, being proportional to the square of
charge, would acquire a factor of four due to such images.
However as the crystal exists only in the semi-space
the deformations only there contribute to the
elastic energy. Therefore the overall factor that arises due to the
presence of the boundary is $4\cdot 1/2=2$:
\begin{equation}
E_{\rm boundary}(r) = 2E_{\rm free}(r) = 
\frac {Yh^2}{2\pi} \ln \left( \frac ra\right).
\end{equation}
This factor can explain the discrepancy between our calculation and the
numerical experiment.

{\em ii) Non-uniform crystal: $N \gg N^*$.} 

In this case as it follows from Eq.~(\ref{n_d}) there are $N_d \sim \sqrt{N}$
dislocations in the island associated with the variations of the electron
density. According to Eq.~(\ref{n_d}) these defects have to be present 
in the bulk of the crystal.
Hence they cannot be arranged into grain boundaries as in the
case considered above. The nearest dislocations repel each other
and one could imagine that they form a crystal themselves.~\cite{Nazarov96} 
We argue however that this cannot be the case. The reason is that
dislocation can occupy a fixed position within an electron lattice cell.
Hence the incommensurability of the electron and dislocation lattices
eventually produces frustration and destroys the long-range order in
the dislocation lattice. Therefore we expect 
that the dislocations form a glassy state.

This implies that the coherence of the crystalline rows in the electron crystal
itself is destroyed.  
We think therefore that in this regime the variations of the total energy
are not periodic.
They retain however the general features described in the introduction
to this section. To understand how these features arise in this
particular case we would like to study the reconstructions in the 
dislocation lattice caused by addition of a new electron. 
To this end it is necessary to consider two cases: when this addition
does not change the total number of defects, and when the number of defects 
is increased. The former case takes place most of the time, while 
the latter happens once in $\sqrt{N}$ electrons when a new 
defect has to appear.

Consider the first case. Electron can be added into the core of dislocation
increasing the length of the extra crystalline row associated with it.
The dislocation after this is displaced by the lattice constant.
Let us find the maximum change in the elastic energy of the island
associated with such a displacement.
It can be expressed through the interaction between dislocations:
\begin{equation}
\delta E_{\rm max} \sim \frac {d^2U(r)} {dr^2} a^2,
\end{equation}
where $U(r)=Yh^2\ln(r)/4\pi $ is the dislocation interaction energy.
The reason why the second derivative is relevant to the calculation
of this energy is that the dislocation is in equilibrium before 
adding the new electron. 
Hence the gradient of the self-energy of the dislocation is zero.
Now we have to remember that there are $\sqrt{N}$ dislocations in
the island. Hence the actual addition energies range from zero to $\delta E_{\rm max}$
with the average level spacing 
\begin{equation}
\delta E \sim \frac {\delta E_{\rm max}} {\sqrt{N}} \sim \frac {e^2 a}{\kappa R^2}.
\end{equation}
This quantity describes the energy one pays when adding an electron at the
fixed number of dislocations. 
This is a correction to the energy spacing. 
Multiplied by the total number of electrons
between two consecutive additions of dislocations $\sqrt{N}$ 
it gives the variation of the chemical potential $\delta E \sqrt{N} \sim e^2/\kappa R$.
This variation should be equal to the discontinuity of the 
chemical potential when a new dislocation is added to the island,
as the average correction to this quantity due to elastic effects does not grow.
Thus the drop of chemical potential is equal to $\delta \mu \sim e^2/\kappa R$.
This is consistent with Eq.~(\ref{delta_Delta}).

Thus elongated crystalline lines draw dislocations from the
center to the periphery of the island. Eventually a new row
has to be inserted in the center. This event manifests itself
in the appearance of a new pair of dislocations. The distance 
from the center $\xi$ at which they are situated can be found from  
Eq.~(\ref{n_d}) by stating that
\begin{equation}
n_d(\xi ) \sim \frac {1}{\xi ^2}.
\end{equation}
This results in
\begin{equation}
\xi \sim a^{1/3} R^{2/3}.
\end{equation}
The energetics of switching between two branches corresponding to different 
number of dislocations has already been discussed above 
[see Eq.~(\ref{delta_Delta})].

%
%
%
%

\section{Conclusions}
\label{Conclusions}

In this section we would like to discuss the degree of universality of our
results. The first question is what happens if confinement potential $V(r)$ is
not parabolic. The answer is almost obvious for the hard-disc
interaction. In this case electrons are added to the crystal on the
equipotential determined by the level of chemical potential. 
Therefore only the shape of
this equipotential and the confinement potential gradient
matter for energy fluctuations. It is obvious that situation is almost
identical for any isotropic confinement $V(r)$. It is also easy to generalize our
calculations for the confinements of oval shapes.

In the latter case the theory developed in 
Section~\ref{Inc_Isl_ATheory} can be applied with a few modifications.
As it was explained the variations of energy are associated with
periodic intersections of the equipotentials with the crystalline rows.
The main contribution to these oscillations comes from the ``coherence'' spots of
the size $\sim \sqrt{R\cdot a}$ tangential to the rows. The oscillations of energy
are sensitive then only to the properties of the equipotential in the 
immediate vicinity to those spots. If it can be well approximated
there by a circle Eq.~(\ref{Delta_Omega}) should hold with
additional phase shifts introduced into the arguments of the 
contributions from different rows. These phase shifts arise due to
the deviation of the global shape of the equipotential from the circle,
and express the incoherence of contributions coming from six different
series of rows.
In addition to that each contribution should 
be rescaled according to $\delta \Omega_0 \propto \sqrt{R_i} T_{\mu}$,
where $R_i$ is the curvature radius of the corresponding equipotential.
We would like to emphasize again that this theory works only
for an island of oval shape. It is not applicable for example 
to a square. All of the curvature radii $R_i$ have to be of the order
of the size of the island. 
In general as the contributions
from different terraces are not coherent anymore, the overall amplitude
of oscillations has to be smaller in the oval case compared to the 
case of rotational symmetry. 

Let us discuss the universality of the results with respect to
the choice of the interaction potential. As we have seen 
both short-range and Coulomb interactions give rise  
to the fluctuations of energy of the same functional shape.
We argue that the other forms of long-range interactions,
logarithmic for instance, bring about similar results. 
Thus the energy of a disorder-free cylinder or disk 
of a type-II superconductor filled with the fluxoid lattice
or a rotating cylindrical vessel of the superfluid helium
as a function of the number of vortices\cite{HeliumArticle}
should experience  oscillations similar to those in Fig.~\ref{fig65}. 
To check this prediction
we have performed a numerical calculation in our model 
with the interaction $U(r)=U_0\ln (1/r)$. It revealed a
quasi-periodic correction to the energy of the same functional 
shape as shown in Fig.~\ref{fig65}.
The amplitude of the correction was consistent with the conclusions of 
Section~\ref{Compr_Isl}. 
The configurations were also very similar to the
observed in the Coulomb case. In particular the correlation between
the position of the center of the confinement relative to the crystal and
the oscillations of the energy is also observed.

In conclusion we have studied the charging spectrum of 
the crystal formed by particles in the
parabolic confinement. We considered two forms of the interactions 
between particles: the short-range and Coulomb interactions.
In the computer simulations employing the genetic algorithm we
have observed the oscillations of the ground state energy which 
have a universal form, independent of the form of interaction.
We attribute these oscillations to the combination of two effects:
periodic additions of new crystalline rows and
hopping of the center of the confinement relative to the crystal.
The hops are separated by addition of $\sim N^{1/2}$ electrons.
These hops make a dramatic difference for the addition spectrum of
the island. In the case of the short-range interaction they make the
charging energy negative, so that $\sim N^{1/4}$ new electrons
enter the island simultaneously. This apparent attraction between electrons
is a result of the confinement polaron effect discussed in 
Ref.~\onlinecite{Us_Phyl_Mag}. In the case of Coulomb interaction
such a hop results in an abrupt $\approx 15\%$ decrease in the charging energy.

%
%
%
%

\section{Acknowledgements}
\label{Acknowledgements}

We thank M.~Fogler, A.~I.~Larkin, J.~R.~Morris, D.~R.~Nelson, and N.~Zhitenev for
stimulating discussions and numerous useful suggestions.
This work is supported by NSF grant DMR-9616880.

\appendix

%
%
%

\section{The Total Disclination Charge of The Island and 
the Euler's Theorem}
\label{App_B}
To prove Eq.~(\ref{Nd}) one has to first consider some 
triangulation of the electron lattice. It is convenient
to consider the triangulation in which every electron
is connected by edges to the nearest neighbors.
In the case of triangular lattice an electron in the bulk 
has six nearest neighbors,
while an electron on the surface of the sample has only four.
For some electrons however this number can be different.
For example, an electron in the core of $\pi /3$ disclination
has only five nearest neighbors (see Fig.~\ref{fig70}). Electrons on the surface 
can have the coordination number equal to three. 
We associate such electrons with the $\pi /3$ disclinations
stuck to the surface.

Let us now use the Euler's theorem. For our case it says:
\begin{equation}
v+f-e=1,
\label{Euler}
\end{equation}
where $v$, $f$, and $e$ are the numbers of vertices, faces, and  
edges contained in the graph formed by our triangulation.
All the vertices of the graph can be divided into
the internal ones $v_i$, belonging to the bulk, and the ones on the surface $v_e$. 
The same can be done for the edges:
\begin{equation}
\begin{array}{c}
{v = v_i+v_e}\\
{e = e_i+e_e}.
\end{array}
\label{surf_bulk_division}
\end{equation}
As all the faces of our figure are triangular by construction
the following relationship, expressing the general balance of edges, is true:
\begin{equation}
3f = e_e+2e_i.
\label{faces}
\end{equation}
Next we can relate the number of edges to the number 
of vertices. 
As it was mentioned above
the bulk vertices are connected to six edges while the
surface ones to four. The exceptions are the cores of
disclinations. They have an anomalous coordination
number. Expressing now the overall balance of edges
one can write:
\begin{equation}
2e = 6v_i - \delta v_i + 4v_e - \delta v_e,
\label{edges}
\end{equation}
where $\delta v_i$ and $\delta v_e$ are total deviations from
the normal coordination numbers for the
internal and external vertices respectively.
One can finally use the obvious fact that 
\begin{equation}
v_e = e_e
\end{equation}
to solve the system of equations (\ref{Euler}) - (\ref{edges})
and to obtain:
\begin{equation}
N_c = \delta v_i + \delta v_e = 6.
\end{equation}
This completes the proof of Eq.~(\ref{Nd}). We would like to notice that
a similar theorem is well known for a triangular crystal on the surface 
of a sphere:~\cite{Morris96,Nelson83}
\begin {equation}
N_c = 12.
\end{equation}

\vspace{-0.2in}

\end{multicols}

\begin{references}
\bibitem{RemarkOnSmallIncrements} For our case the theorem of small
increments can be proved in the following way: 
$\delta E(N) \equiv E(N)-\bar{E}(N)=\Omega (\bar{\mu}(N)+\delta \mu (N)) - %
\bar{\Omega}(\bar{\mu}(N)) + N\delta \mu (N) \approx %
\Omega (\bar{\mu}(N)) + \delta \mu (N) d\Omega /d\mu - \bar{\Omega}(\bar{\mu}(N))%
+ N\delta \mu (N)$. Now we use the identity $d\Omega /d\mu = -N$ to obtain
$\delta E(N) = \Omega (\bar{\mu}(N)) - \bar{\Omega}(\bar{\mu}(N)) \equiv %
\delta \Omega(\bar{\mu}(N))$
\end{references}
\end{document}